\documentclass[11pt,final,onecolumn]{IEEEtran}

\PassOptionsToPackage{}{tocbibind}
\PassOptionsToPackage{style=ieee}{biblatex}
\usepackage[colorlinks,tables=false,bibtex=true]{preamble}
\RequirePackage{upgreek}
\graphicspath{{/}}

\makeatletter
	\ifbool{@bibtex}{}{
		\addbibresource{IEEEabrv,UWA,PaperList,ownpub}
	}
\makeatother

\newcommand{\Gangle}{\mathcal{G}_{\mathrm{A}}}
\newcommand{\Kzero}{\mathcal{K}_{\vec{0}}}
\newcommand{\Kone}{\mathcal{K}_{\vec{1}}}
\newcommand{\Kb}{\mathcal{K}_{\vec{b}}}

\DeclareMathOperator{\IdR}{\mathrm{Id}_{\mathrm{R}}}

\newcommand{\pref}[1]{(\hyperref[#1]{C\ref{#1}})}
\newcounter{asmplistctr}

\title{Fundamental Limits of Blind Deconvolution\\\partname~II: Sparsity-Ambiguity Trade-offs}
\author{Sunav~Choudhary,~\IEEEmembership{Student~Member,~IEEE,} and~Urbashi~Mitra,~\IEEEmembership{Fellow,~IEEE}%
		\thanks{This work has been funded in part by the following grants and organizations:~ONR~N00014-09-1-0700, AFOSR~FA9550-12-1-0215, NSF~CNS-0832186, NSF~CNS-1213128 and NSF~CCF-1117896.
		Parts of this paper were presented at the IEEE International Symposium on Information Theory (ISIT), Honolulu, Hawai'i, June 29 - July 4, 2014~\cite{choudhary2014sbdidentifiability} and at the IEEE Global Conference on Signal and Information Processing (GlobalSIP), Atlanta, Georgia, Dec. 3-5, 2014~\cite{choudhary2014subspaceblind}.}%
		\thanks{S.~Choudhary and U.~Mitra are with the Ming Hsieh Department of Electrical Engineering, Viterbi School of Engineering, University of Southern California, Los Angeles CA 90089, USA (email: \href{mailto:sunavcho@usc.edu}{\protect\nolinkurl{sunavcho@usc.edu}}, \href{mailto:ubli@usc.edu}{\protect\nolinkurl{ubli@usc.edu}})}}

\begin{document}
	\maketitle

	\begin{abstract}
		Blind deconvolution is an ubiquitous non-linear inverse problem in applications like wireless communications and image processing.
		This problem is generally ill-posed since signal identifiability is a key concern, and there have been efforts to use sparse models for regularizing blind deconvolution to promote signal identifiability.
		\partname~I of this two-part paper establishes a measure theoretically tight characterization of the ambiguity space for blind deconvolution and unidentifiability of this inverse problem under unconstrained inputs.
		\partname~II of this paper analyzes the identifiability of the canonical-sparse blind deconvolution problem and establishes surprisingly strong negative results on the sparsity-ambiguity trade-off scaling laws.
		Specifically, the ill-posedness of canonical-sparse blind deconvolution is quantified by exemplifying the dimension of the unidentifiable signal sets.
		An important conclusion of the paper is that canonical sparsity occurring naturally in applications is insufficient and that coding is necessary for signal identifiability in blind deconvolution.
		The methods developed herein are applied to a second-hop channel estimation problem to show that a family of subspace coded signals (including repetition coding and geometrically decaying signals) are unidentifiable under blind deconvolution.
	\end{abstract}

	\IEEEpeerreviewmaketitle
	
	\section{Introduction}
		\label{sec:intro}
		\IEEEPARstart{B}{lind} deconvolution is a challenging open problem in signal processing that manifests itself in a host of applications like image de-blurring, blind system identification, and blind equalization.
		The difficulty of blind deconvolution stems from its ill-posedness in the sense of Hadamard~\cite{Hadamard1902,Tikhonov1986} as well as non-linearity of the problem.
		In practice, additional application specific prior information on the unknown signals is necessary to render this inverse problem better behaved.
		\emph{Sparsity} based models have been used extensively in the past decade to capture hidden signal structures in many applications of interest.
		While there have been a few attempts at exploiting sparsity based priors for blind deconvolution type of problems~\cite{ahmed2012blind,kammoun2010robustness,herrity2008blind,barchiesi2011dictionary,hegde2011sampling,grady2005survey}, strong theoretical guarantees on signal identifiability for such sparse models are largely missing with the exception of \cite{ahmed2012blind} (which proves an interesting positive result under idealized assumptions).
		Exploring more realistic sparsity assumptions pertaining to multi-hop channel estimation, we prove some surprising negative results for signal identifiability in the present article.
		\partname~I of this paper~\cite{choudhary2014limitsBDambiguity} was devoted to the characterization of the ambiguity space associated with blind deconvolution and the proof of a strong unidentifiability result.
		\partname~II of the paper utilizes the results in \partname~I to characterize the ambiguity space for sparsity constrained blind deconvolution problems.
		In particular, canonical-sparse blind deconvolution is considered and a trade-off is derived between the level of sparsity in the unknown signals and the dimension of the ambiguity space.
		Unfortunately, the derived scaling laws turn out to be unfavorable when compared to analogous results in compressed sensing~\cite{donoho2006compressed}.
		Furthermore, an application of the methodology to a theoretical abstraction of the second-hop channel estimation problem, yields an unfavorable sparsity-ambiguity trade-off for a family of subspace coded signals that include repetition coding and geometrically decaying signals.

		\subsection{Related Work}
			\label{sec:prior art}
			We briefly reviewed prior research on the multi-channel version of blind deconvolution~\cite{meraim1997blind,johnson1998blind,liu1996recentblind} in \partname~I (see~\cite{choudhary2014limitsBDambiguity} for details).
			We briefly recap prior art discussed in \partname~I of the paper that pertains specifically to blind deconvolution in single-input-single-output (SISO) systems, before mentioning more related research.
			
			For SISO blind deconvolution (\ie~blind channel estimation),~\cite{manton2003totallyblind} showed that input signal realizations with no inter-symbol-interference (ISI) could be uniquely recovered up to scalar multiplicative ambiguities.
			In contrast, we showed in \partname~I of the paper that if ISI were allowed, blind deconvolution is ill-posed to the point of almost every input signal pair being unidentifiable.
			In the papers~\cite{asif2009random,ahmed2012blind}, blind deconvolution was cast as a rank one matrix recovery problem for the development of convex programming based computational heuristics and a proof of correctness was provided assuming a random subspace precoding with sufficient redundancy.
			We adopted this framework in our earlier works~\cite{choudhary2012onidentifiability,choudhary2012identifiabilitybounds,choudhary2013bilinear} on an information theoretic characterization of identifiability of the solution to general bilinear inverse problems and showed good predictive performance of the theory for blind deconvolution with Gaussian/Rademacher inputs.
			Whereas~\cite{choudhary2013bilinear} derived conditions for identifiability (analogous to achievability results in information theory) and developed an abstract theory of regularized bilinear inverse problems, making assumptions no more specific than non-convex cone constrained priors, the present paper makes a significant departure in the sense that more explicit sparsity and subspace priors are considered that are motivated by applications in multi-hop channel estimation~\cite{richard2008sparse,choudhary2012sparse,michelusi2011HSD} and the focus is on characterizing the inherent \textit{unidentifiability} of blind deconvolution (akin to converse/impossibility results in information theory).
			We point out that beside multi-hop communications, convolution with sparse vectors is an intuitive modeling assumption in other applications as well, \eg~high resolution astronomical imaging~\cite{hegde2011sampling}, modeling echoing audio signals~\cite{barchiesi2013LearningIncoherentDictionaries}, ultrasound imaging~\cite{tur2011Innovationratesampling}, and sporadic communications in 5G systems~\cite{jung2014SparseModelUncertainties}.

			A promising identifiability analysis was proposed in~\cite{kammoun2010robustness}, leveraging results from~\cite{gribonval2010dictionary} on matrix factorization for sparse dictionary learning using the \lonenorm{} and $\ell_q$ quasi-norm for $0 < q < 1$.
			Their approach and formulation differ from ours in two important aspects.
			Firstly, we are interested in single-in-single-out (SISO) systems whereas~\cite{kammoun2010robustness} deals with single-in-multiple-out (SIMO) systems.
			Secondly,~\cite{kammoun2010robustness} analyzes identifiability as a \textit{local} optimum to a \textit{non-convex} $\ell_1$ (or $\ell_q$ for $0 < q < 1$) optimization and hence is heavily dependent on the algorithmic formulation, whereas we consider the solution as the local/global optimum to the $\ell_0$ optimization problem and our impossibility results are information theoretic in nature, implying that they hold regardless of algorithmic formulation.
			We emphasize that the constrained $\ell_{1}$ optimization formulation in \cite{kammoun2010robustness} is non-convex and therefore does not imply existence of provably correct and efficient recovery algorithms, despite identifiability of the channel.
			Although it would be interesting to try and extend their approach to SISO systems and compare with our results, this is non-trivial and beyond the scope of the present paper.

			An inverse problem closely related to blind deconvolution is the Fourier phase retrieval problem~\cite{jaganathan2013phase,fannjiang2012phase,candes2013phase} where a signal has to be reconstructed from its autocorrelation function.
			This is clearly a special case of the blind deconvolution problem with much fewer degrees of freedom and allows identifiability and tractable recovery with a sparsity prior on the signal~\cite{jaganathan2013phase}.
			A second important difference is that after \emph{lifting}~\cite{balas2005projection}, the Fourier phase retrieval problem has one linear constraint involving a positive semidefinite matrix.
			This characteristic is known to be helpful in the conditioning of the inverse problem and in the development of recovery algorithms~\cite{beck2009matrixQP}.
			While the blind deconvolution problem does not enjoy the same advantage, this approach seems to be a good avenue to explore if additional constraints are allowed.
			This is a future direction of research.

			As in \partname~I, we rely on the \emph{parameter counting} heuristic to quantify `dimension' of a set/space.
			The parameter counting heuristic can be made rigorous using the framework of \emph{Hausdorff dimension}~\cite{mattila1995geometryofsets}, but we shall not bridge this gap in the present paper owing to space limitations.

		\subsection{Contributions and Organization}
			\label{sec:contributions}
			In this work, we quantify unidentifiability for certain families of noiseless sparse blind deconvolution problems under a \emph{non-asymptotic} and \emph{non-statistical} setup.
			Specifically, given model orders $m,n \in \setZ_{+}$, we investigate the trade-off between the level of sparsity in the unknown signal pair $\bb{\vec{x}_{\ast}, \vec{y}_{\ast}} \in \mathcal{K} \subseteq \setR^{m} \times \setR^{n}$ and the dimension of the unidentifiable subset of signals in $\mathcal{K}$.
			While \partname~I of the paper focused on developing a description of the ambiguity space, \partname~II studies the interaction of this ambiguity space with canonical-sparse vectors and further extends the techniques to consider repetition coded and geometrically decaying vectors.
			To motivate our signal choices, the specific application we consider is multi-hop sparse channel estimation for relay assisted communication described in \sectionname~\ref{sec:results with coding}.
			Like \partname~I, our focus is on an \emph{algorithm independent} identifiability analysis and hence we shall \emph{not} examine efficient/polynomial-time algorithms, but rather show information theoretic impossibility results.
			Our approach leads to the following novelties for \partname~II (refer to~\cite{choudhary2014limitsBDambiguity} for contributions of \partname~I of the paper).

			\begin{enumerate}
				\item	We show that sparsity in the canonical basis is not sufficient to ensure identifiability, even in the presence of perfect model order information, and we construct non-zero dimensional unidentifiable subsets of the domain $\mathcal{K}$ for any given support set of $\vec{x}_{\ast}$ (or $\vec{y}_{\ast}$) as evidence to quantify the sparsity-ambiguity trade-off.
						This is the content of \theoremname~\ref{thm:sparse unident moderate} in \sectionname~\ref{sec:sparse deconv}.
				\item	We consider a theoretical abstraction to a multi-hop channel estimation problem and extend our unidentifiability results to this setting.
						Specifically, we show that other types of side information like repetition coding or geometric decay for the unknown vectors are still insufficient for identifiability, and exhibit an ambiguity trade-off analogous to the case of canonical-sparsity.
						This is the content of \theoremname~\ref{thm:sparse cooperative coding unident} and \corollariesname~\ref{cor:repetition coding unident} and~\ref{cor:sparse geometric unident} in \sectionname~\ref{sec:results with coding}.
			\end{enumerate}

			The rest of the paper is organized as follows.
			\sectionname~\ref{sec:overview} presents constructive numerical examples illustrating the nature of unidentifiable inputs for blind deconvolution.
			\sectionname~\ref{sec:model} recaps the system model, the notion of identifiability, and the lifted reformulation of the blind deconvolution problem from \partname~I of the paper.
			\sectionname~\ref{sec:results without coding} presents the key unidentifiability results for canonical-sparse blind deconvolution and contrasts them against results from \partname~I of the paper.
			\sectionname~\ref{sec:results with coding} extends the techniques and results in \sectionname~\ref{sec:results without coding} to more general families of subspace based structural priors, including repetition coding and geometrically decaying canonical representations.
			\sectionname~\ref{sec:conclusion} concludes the paper.
			Detailed proofs of the major results in the paper appear in \appendicesname~\ref{sec:sparse unident moderate proof}-\ref{sec:sparse cooperative coding unident proof}.

		\subsection{Notational Conventions}
			\label{sec:notation}
			We shall closely follow the notation used in \partname~I of the paper which is recapped below.
			All vectors are assumed to be column vectors unless stated otherwise and denoted by lowercase boldface alphabets~(\eg~$\vec{a}$).
			Matrices are denoted by uppercase boldface alphabets~(\eg~$\mat{A}$).
			The MATLAB\textsuperscript{\circledR} indexing rules are used to denote parts of a vector~(\eg~$\vec{a}\bb{4:6}$ denotes the sub-vector of $\vec{a}$ formed by the $4^{\thp}$, $5^{\thp}$ and $6^{\thp}$ elements of $\vec{a}$).
			The all zero vector/matrix (respectively all one vector) shall be denoted by $\vec{0}$ (respectively $\vec{1}$) and its dimension would be clear from the usage context.
			For vectors and/or matrices, $\tpose{\bb{\cdot}}$ and $\rank{\cdot}$ respectively return the transpose and rank of their argument, whenever applicable.
			Special sets are denoted by uppercase blackboard bold font~(\eg~$\setR$ for real numbers) and a subscripted `$+$' sign would denote the non-negative subset whenever applicable~(\eg~$\setR_{+}$ for non-negative real numbers).
			Other sets are denoted by uppercase calligraphic font~(\eg~$\mathcal{S}$).
			For any set $\mathcal{S}$, $\card{\mathcal{S}}$ shall denote its cardinality.
			Linear operators on matrices are denoted by uppercase script font~(\eg~$\mathscr{S}$).

			To avoid unnecessarily heavy notation, we shall adopt the following convention: The scope of both vector variables (like $\vec{x}$, $\vec{y}$, $\vec{u}$, $\vec{v}$, \etc) as well as matrix variables (like $\mat{X}$, $\mat{Y}$, \etc) are restricted to individual theorems and/or proofs.
			Their meanings are allowed to differ \emph{across} theorems and proofs (and even across disjoint subparts of the same proof when there is no risk of confusion), thus facilitating the reuse of variable names across different theorems and avoiding heavy notation.

	\section{Background}
		\label{sec:background}
		In this section, we review the system model from \partname~I of the paper and overview a new result from \partname~II of the paper with a numerical example.
		The results in \partname~II of the paper continue to use the same system model, but with a different feasible set of signals.

		\subsection{System Model}
			\label{sec:model}
			We consider the noiseless linear convolution observation model
			\begin{equation}
				\vec{z} = \vec{x} \star \vec{y},
				\label{eqn:model}
			\end{equation}
			where $\star \colon \setR^{m} \times \setR^{n} \to \setR^{m + n - 1}$ denotes the linear convolution map, $\bb{\vec{x}, \vec{y}} \in \mathcal{K} \subseteq \setR^{m} \times \setR^{n}$ denotes the pair of input signals from a restricted domain $\mathcal{K}$, and $\vec{z} \in \setR^{m+n-1}$ is the vector of observations given by
			\begin{equation}
				\vec{z}\bb{l} =	\begin{cases}
									{\displaystyle \sum_{j = 1}^{\min\bb{l,m}} \vec{x}\bb{j} \vec{y}\bb{l+1-j}},
									&	1 \leq l \leq n,	\\
									{\displaystyle \sum_{j = l+1-n}^{\min\bb{l,m}} \vec{x}\bb{j} \vec{y}\bb{l+1-j}},
									&	1 \leq l-n \leq m-1.
								\end{cases}
				\label{eqn:conv defn}
			\end{equation}
			The corresponding inverse problem of constrained blind linear deconvolution is to find the vector pair $\bb{\vec{x}, \vec{y}} \in \mathcal{K}$ given the noiseless observation $\vec{z}$ and is symbolically represented by the feasibility problem
			\find{\bb{\vec{x}, \vec{y}}}
			{\vec{x} \star \vec{y} = \vec{z}, \sep \bb{\vec{x}, \vec{y}} \in \mathcal{K}.}
			{\label{prob:find_xy}}
			Assuming that the model orders $m$ and $n$, respectively, of vectors $\vec{x}$ and $\vec{y}$ are fixed and known \textit{a priori}, we are concerned with whether \problemname~\eqref{prob:find_xy} admits a unique solution.
			Since bilinearity of the convolution operator implies the inherent scaling ambiguity in \problemname~\eqref{prob:find_xy} given by
			\begin{equation}
				\vec{x} \star \vec{y} = \alpha \vec{x} \star \frac{1}{\alpha} \vec{y}, \quad \forall \alpha \neq 0,
				\label{eqn:scaling ambiguity}
			\end{equation}
			the question of uniqueness only makes sense modulo such scalar multiplicative factors.
			This leads us to the definition of identifiability in \partname~I that we restate below.

			\begin{definition}[Identifiability]
				\label{defn:identifiability}
				A vector pair $\bb{\vec{x}, \vec{y}} \in \mathcal{K} \subseteq \setR^{m} \times \setR^{n}$ is identifiable within $\mathcal{K}$ with respect to the linear convolution map~$\star$, if $\forall \bb{\vec{x}', \vec{y}'} \in \mathcal{K}$ satisfying $\vec{x} \star \vec{y} = \vec{x}' \star \vec{y}'$, $\exists \alpha \neq 0$ such that $\bb{\vec{x}', \vec{y}'} = \bb{\alpha \vec{x}, \frac{1}{\alpha} \vec{y}}$.
			\end{definition}

			It is easy to see that \definitionname~\ref{defn:identifiability} induces an equivalence structure on the set of identifiable pairs in $\mathcal{K}$.
			For future reference, we define the equivalence relation $\IdR \fcolon \mathcal{K} \times \mathcal{K} \to \cc{0,1}$ as follows.
			Given any $\bb{\vec{x}, \vec{y}}, \bb{\vec{x}', \vec{y}'} \in \mathcal{K}$, $\IdR \bb[\big]{\bb{\vec{x}, \vec{y}}, \bb{\vec{x}', \vec{y}'}} = 1$ if and only if $\exists \alpha \neq 0$ such that $\bb{\vec{x}', \vec{y}'} = \bb{\alpha \vec{x}, \frac{1}{\alpha} \vec{y}}$.
			It is straightforward to check that $\IdR\bb{\cdot,\cdot}$ is indeed an equivalence relation.
			Let $\mathcal{K}/\IdR$ denote the set of equivalence classes of $\mathcal{K}$ induced by $\IdR\bb{\cdot,\cdot}$, and for any $\bb{\vec{x}, \vec{y}} \in \mathcal{K}$ let $\BB{\bb{\vec{x}, \vec{y}}} \in \mathcal{K}/\IdR$ denote the equivalence class containing $\bb{\vec{x}, \vec{y}}$.
			Then \definitionname~\ref{defn:identifiability} amounts to declaring a vector pair $\bb{\vec{x}, \vec{y}} \in \mathcal{K}$ as identifiable if and only if every $\bb{\vec{x}', \vec{y}'} \in \mathcal{K}$ with $\BB{\bb{\vec{x}', \vec{y}'}} \neq \BB{\bb{\vec{x}, \vec{y}}}$ satisfies $\vec{x} \star \vec{y} \neq \vec{x}' \star \vec{y}'$.

			Using the \emph{lifting} technique from optimization~\cite{balas2005projection}, \problemname~\eqref{prob:find_xy} can be reformulated as a rank minimization problem subject to linear equality constraints~\cite{choudhary2012onidentifiability,asif2009random}
			\minimize{\mat{W}}
			{\rank{\mat{W}}}
			{\mathscr{S}\bb{\mat{W}} = \vec{z}, \sep \mat{W} \in \mathcal{W},}
			{\label{prob:rank}}
			where $\mathcal{W} \subseteq \setR^{m \times n}$ is \emph{any} set satisfying
			\begin{equation}
				\mathcal{W} \bigcap \set{\mat{W} \in \setR^{m \times n}}{\rank{\mat{W}} \leq 1}
				= \set{\vec{x} \tpose{\vec{y}}}{\bb{\vec{x}, \vec{y}} \in \mathcal{K}},
				\label{eqn:set change}
			\end{equation}
			and $\mathscr{S} \colon \setR^{m \times n} \to \setR^{m+n-1}$ is the unique linear operator (henceforth referred to as the \emph{lifted linear convolution operator}) satisfying
			\begin{equation}
				\mathscr{S}\bb{\vec{x} \tpose{\vec{y}}} = \vec{x} \star \vec{y}, \quad \forall \bb{\vec{x}, \vec{y}} \in \setR^{m} \times \setR^{n}
				\label{eqn:lifted op}
			\end{equation}
			and admitting an explicit closed form specification (see \partname~I~\cite{choudhary2014limitsBDambiguity}, \remarkname~1).
			\problemsname~\eqref{prob:find_xy} and~\eqref{prob:rank} are equivalent for $\vec{z} \neq \vec{0}$ (see~\cite{choudhary2013bilinear} for details).
			Hence, owing to the analytical simplicity of \problemname~\eqref{prob:rank}, identifiability of the solution to \problemname~\eqref{prob:find_xy} is derived using the properties of \problemname~\eqref{prob:rank}.
			By construction, the optimal solution to \problemname~\eqref{prob:rank} is a rank one matrix $\mat{W}_{\opt}$ and its singular value decomposition $\mat{W}_{\opt} = \sigma_{\opt} \vec{u}_{\opt} \tpose{\vec{v}_{\opt}}$ yields a solution $\bb{\vec{x}, \vec{y}}_{\opt} = \bb{\sqrt{\sigma_{\opt}} \vec{u}_{\opt}, \sqrt{\sigma_{\opt}} \vec{v}_{\opt}}$ to \problemname~\eqref{prob:find_xy}.

		\subsection{An Overview of the Results}
			\label{sec:overview}
			Let us interpret the role of $\mathcal{K} \subseteq \setR^{m} \times \setR^{n}$ for the discrete-time blind linear deconvolution problem~\eqref{prob:find_xy}.
			Suppose that $\bb{\vec{x}_{\ast}, \vec{y}_{\ast}} \in \mathcal{K}$ is the ground truth resulting in the observation $\vec{z} = \vec{x}_{\ast} \star \vec{y}_{\ast}$.
			The set $\mathcal{K}$ captures application specific constraints on the signal pair $\bb{\vec{x}_{\ast}, \vec{y}_{\ast}}$ (in our case, sparsity constraints on the true solution $\bb{\vec{x}_{\ast}, \vec{y}_{\ast}}$) by restricting the feasible set of signal choices in \problemname~\eqref{prob:find_xy}.
			Without such a restriction, the solution to \problemname~\eqref{prob:find_xy} would most likely be unidentifiable (infact, \theoremname~2 in \partname~I shows that almost every ground truth $\bb{\vec{x}_{\ast}, \vec{y}_{\ast}}$ \wrt~the Lebesgue measure is unidentifiable in the absence of constraints).
			In the sequel, we solely consider $\mathcal{K}$ to be a separable cone in $\setR^{m} \times \setR^{n}$, \ie~there exist sets $\mathcal{D}_{1} \subseteq \setR^{m}$ and $\mathcal{D}_{2} \subseteq \setR^{n}$ such that $\mathcal{K} = \mathcal{D}_{1} \times \mathcal{D}_{2}$ and for every $\alpha > 0$, $\bb{\vec{x}, \vec{y}} \in \mathcal{K}$ implies that $\bb{\alpha \vec{x}, \alpha \vec{y}} \in \mathcal{K}$.
			Such a separability of $\mathcal{K}$ is motivated by the observation that in many applications of interest $\vec{x}_{\ast}$ and $\vec{y}_{\ast}$ are unrelated (like $\vec{x}_{\ast}$ and $\vec{y}_{\ast}$ may respectively represent the unknown source and the unknown channel in a blind channel estimation problem).

			\theoremname~\ref{thm:sparse cooperative coding unident} in \sectionname~\ref{sec:partial cooperation} is the most general result of this part of the paper.
			However, we shall not state it here since it requires some technical definitions that would appear unmotivated at this stage.
			Instead, we shall succinctly overview the results pertaining to the special case of canonical-sparse blind deconvolution.
			We revisit the numerical example from \partname~I of the paper and examine its identifiability (according to \definitionname~\ref{defn:identifiability}) under discrete-time blind linear deconvolution with canonical-sparsity constraints.
			In preparation for the result, we first define a parameterized family of canonical-sparse cones for an arbitrary integer $d \geq 3$ and any index set $\Lambda \subseteq \cc{2,3,\dotsc,d-1}$ as
			\begin{equation}
				\Kzero\bb{\Lambda,d}	\triangleq \set{\vec{w} \in \setR^{d}}{\vec{w}\bb{1} \neq 0, \vec{w}\bb{d} \neq 0, \vec{w}\bb{\Lambda} = \vec{0}},
				\label{eqn:sparse domain}
			\end{equation}
			\ie~$\Lambda$ denotes the set of indices that are \emph{zero} across \emph{all} vectors in $\Kzero\bb{\Lambda,d}$.
			Secondly, let us denote the Minkowski sum of the sets $\cc{-1}$ and $\Lambda$, by the shorthand notation $\Lambda - 1$, defined as
			\begin{equation}
				\Lambda - 1 = \cc{-1} + \Lambda \triangleq \set{j - 1}{j \in \Lambda}.
			\end{equation}
			Jumping ahead to provide a concrete example, the following unidentifiability result can be obtained as a corollary to \theoremname~\ref{thm:sparse unident moderate} in \sectionname~\ref{sec:sparse deconv}.

			\begin{corollary}
				\label{cor:mixed unident moderate}
				Let $m \geq 5$ and $n \geq 2$ be arbitrary integers and $\emptyset \neq \Lambda \subseteq \cc{3,4,\dotsc,m-2}$ denote a set of indices.
				Let $\mathcal{K} = \Kzero\bb{\Lambda,m} \times \setR^{n}$ be the structured feasible set in \problemname~\eqref{prob:find_xy} and define $p \triangleq \card{\Lambda \bigcup \bb{\Lambda - 1}}$.
				Then there exists a set $\mathcal{G}_{\ast} \subseteq \mathcal{K}/\IdR$ of dimension $\bb{m+n-p-1}$ such that every signal pair $\bb{\vec{x}, \vec{y}} \in \mathcal{G}_{\ast}$ is unidentifiable by \definitionname~\ref{defn:identifiability}.
			\end{corollary}

			\corollaryname~\ref{cor:mixed unident moderate} cannot be checked in its full generality by considering examples, since it is a statement about an uncountably infinite number of vectors.
			However, we will re-analyze the numerical example in \partname~I of the paper (within canonical-sparse feasible sets) to capture the ideas behind \corollaryname~\ref{cor:mixed unident moderate}.
			As in \partname~I, consider the vectors
			\begin{subequations}
				\begin{alignat}{2}
					\vec{x}_{1}	& = \tpose{\bb{1,0,1,0,0,0,0,0,1,0,1}},	& \quad \vec{y}_{1}	& = \tpose{\bb{1,0,0,0,1,0,0}}, \\
					\vec{x}_{2}	& = \tpose{\bb{1,0,0,0,0,0,0,0,1,0,0}},	& \quad \vec{y}_{2}	& = \tpose{\bb{1,0,1,0,1,0,1}},
				\end{alignat}
			\end{subequations}
			resulting in
			\begin{equation}
				\vec{x}_{1} \star \vec{y}_{1} = \vec{x}_{2} \star \vec{y}_{2} = \tpose{\bb{1,0,1,0,1,0,1,0,1,0,1,0,1,0,1,0,0}}
				\label{eqn:common convolved output}
			\end{equation}
			with $\vec{x}_{1}$ and $\vec{x}_{2}$ being non-collinear.
			Clearly, the pairs $\bb{\vec{x}_{1}, \vec{y}_{1}}$ and $\bb{\vec{x}_{2}, \vec{y}_{2}}$ are unidentifiable within the domain $\setR^{11} \times \setR^{7}$.
			Setting $d = 11$ and $\Lambda = \cc{4,5,6,7,8}$ in \eqref{eqn:sparse domain} gives $\vec{x}_{1}, \vec{x}_{2} \in \Kzero\bb{\Lambda,11}$ and therefore $\bb{\vec{x}_{1}, \vec{y}_{1}}, \bb{\vec{x}_{2}, \vec{y}_{2}} \in \Kzero\bb{\Lambda,11} \times \setR^{7}$ are still unidentifiable within $\mathcal{K} = \Kzero\bb{\Lambda,11} \times \setR^{7}$.

			Next, we show that the \emph{rotational ambiguity} over $\setR^{11} \times \setR^{7}$ (discussed in \partname~I) stays valid over $\Kzero\bb{\Lambda,11} \times \setR^{7}$ and leads to an uncountable set of unidentifiable pairs in $\mathcal{K}$.
			Let $\vec{z}_{0} = \vec{x}_{1} \star \vec{y}_{1} = \vec{x}_{2} \star \vec{y}_{2}$ denote the common convolved output in \eqref{eqn:common convolved output} and consider the parameterized vectors
			\begin{subequations}
				\label{eqn:rotational transform}
				\begin{alignat}{2}
					\vec{x}_{1}'	& = \vec{x}_{1} \cos \theta - \vec{x}_{2} \sin \theta,	& \quad \vec{y}_{1}'	& = \vec{y}_{1} \sin \phi - \vec{y}_{2} \cos \phi,	\\
					\vec{x}_{2}'	& = \vec{x}_{1} \cos \phi - \vec{x}_{2} \sin \phi,	& \quad \vec{y}_{2}'	& = \vec{y}_{1} \sin \theta - \vec{y}_{2} \cos \theta,
				\end{alignat}
			\end{subequations}
			where $\theta \neq \phi$ are the parameters, and $\cc{\vec{x}_{1}, \vec{x}_{2}, \vec{y}_{1}, \vec{y}_{2}}$ acts as the set of seed vectors for the above transformation.
			Clearly, $\theta \neq \phi$ and non-collinearity of $\vec{x}_{1}$ and $\vec{x}_{2}$ imply that $\vec{x}_{1}'$ and $\vec{x}_{2}'$ are linearly independent.
			A simple algebraic manipulation reveals that
			\makeatletter
				\if@twocolumn
					\begin{equation}
						\begin{split}
							\MoveEqLeft \vec{z}_{0} \sin \bb{\theta + \phi} - \vec{x}_{2} \star \vec{y}_{1} \sin \theta \sin \phi - \vec{x}_{1} \star \vec{y}_{2} \cos \theta \cos \phi	\\
							& = \vec{x}_{1}' \star \vec{y}_{1}' = \vec{x}_{2}' \star \vec{y}_{2}',
						\end{split}
						\label{eqn:parametrized example cos sin}
					\end{equation}
				\else
					\begin{equation}
						\vec{x}_{1}' \star \vec{y}_{1}' = \vec{x}_{2}' \star \vec{y}_{2}' = \vec{z}_{0} \sin \bb{\theta + \phi} - \vec{x}_{2} \star \vec{y}_{1} \sin \theta \sin \phi - \vec{x}_{1} \star \vec{y}_{2} \cos \theta \cos \phi,
						\label{eqn:parametrized example cos sin}
					\end{equation}
				\fi
			\makeatother
			rendering both $\bb{\vec{x}_{1}', \vec{y}_{1}'}$ and $\bb{\vec{x}_{2}', \vec{y}_{2}'}$ unidentifiable within $\setR^{11} \times \setR^{7}$.
			Since $\vec{x}_{1}, \vec{x}_{2} \in \Kzero\bb{\Lambda,11}$, \eqref{eqn:rotational transform} implies that $\vec{x}_{1}', \vec{x}_{2}' \in \Kzero\bb{\Lambda,11}$, thus rendering $\bb{\vec{x}_{1}', \vec{y}_{1}'},\bb{\vec{x}_{2}', \vec{y}_{2}'} \in \Kzero\bb{\Lambda,11} \times \setR^{7}$ unidentifiable within $\mathcal{K} = \Kzero\bb{\Lambda,11} \times \setR^{7}$.
			Since $\bb{\theta, \phi} \in [0, \pi)^{2}$ describes a two dimensional parameter space, \eqref{eqn:rotational transform} and \eqref{eqn:parametrized example cos sin} imply that the unidentifiable subset of $\Kzero\bb{\Lambda,11} \times \setR^{7}$ is at least two dimensional and therefore hints towards \corollaryname~\ref{cor:mixed unident moderate}.
			Thus, having a sparse support does not help with identifiability here.

	\section{Unidentifiability under Canonical Sparsity}
		\label{sec:results without coding}
		We shall use \emph{identifiability} in the sense of \definitionname~\ref{defn:identifiability}.
		In \sectionname~\ref{sec:supporting lemmas}, we recap the partially parametric characterization of the ambiguity space of unconstrained blind deconvolution (\lemmaname~\ref{lem:rank-2 nullspace}), the pathological cases for unidentifiability, and a rotation based representation cum decomposition result (\lemmaname~\ref{lem:finite quotient set}) discussed in \partname~I of the paper.
		In \sectionname~\ref{sec:sparse deconv}, we state our main unidentifiability result for canonical-sparse blind deconvolution as \theoremname~\ref{thm:sparse unident moderate} and contrast it with the almost everywhere unidentifiability result for non-sparse blind deconvolution from \partname~I of the paper.
		\sectionname~\ref{sec:mixed extensions} states a stronger result for the feasible set in \corollaryname~\ref{cor:mixed unident moderate}.
		Throughout this section, we assume that $\mathcal{K}$ represents the (not necessarily convex) feasible \emph{cone} in \problemname~\eqref{prob:find_xy}, \ie~$\forall \bb{\vec{x}, \vec{y}} \in \mathcal{K}$ one has $\bb{\alpha\vec{x}, \alpha\vec{y}} \in \mathcal{K}$ for every $\alpha \neq 0$, but $\mathcal{K}$ is allowed to change from theorem to theorem.

		\subsection{Representation Lemmas and Pathological Cases}
			\label{sec:supporting lemmas}
			Let $\mathscr{S}\bb{\cdot}$ denote the lifted linear convolution operator as described in \sectionname~\ref{sec:model}.
			We denote the rank-$k$ null space of $\mathscr{S}\bb{\cdot}$ by $\mathcal{N}\bb{\mathscr{S}, k}$ and define it as
			\begin{equation}
				\mathcal{N}\bb{\mathscr{S}, k} \triangleq \set{\mat{Q} \in \setR^{m \times n}}{\rank{\mat{Q}} \leq k,\, \mathscr{S}\bb{\mat{Q}} = \vec{0}}.
			\end{equation}
			We note that the rank one null space of $\mathscr{S}\bb{\cdot}$ is trivial, \ie~$\mathcal{N}\bb{\mathscr{S}, 1} = \cc{\mat{0}}$, and this property of linear convolution is implicitly used to prove equivalence of \problemsname~\eqref{prob:find_xy} and~\eqref{prob:rank} for $\vec{z} \neq \vec{0}$ in~\cite{choudhary2013bilinear}.
			$\mathcal{N}\bb{\mathscr{S}, 1} = \cc{\mat{0}}$ follows from interpreting convolution as polynomial multiplication, since the product of two real polynomials is identically zero if and only if at least one of them is identically zero.
			The following lemma (borrowed from \partname~I of this paper) describes a subset of $\mathcal{N}\bb{\mathscr{S}, 2}$, the rank two null space of $\mathscr{S}\bb{\cdot}$, and is used in the proofs of the results in the sequel.

			\begin{lemma}[from~\cite{choudhary2014limitsBDambiguity}]
				\label{lem:rank-2 nullspace}
				Let $m,n \geq 2$ and $\mat{Q} \in \setR^{m \times n}$ admit a factorization of the form
				\begin{equation}
					\mat{Q} =	\begin{bmatrix}
									\vec{u}	&	0	\\
									0	&	-\vec{u}
								\end{bmatrix}
								\begin{bmatrix}
									0	&	\tpose{\vec{v}}	\\
									\tpose{\vec{v}}	&	0
								\end{bmatrix},
					\label{eqn:rank-2 nullspace}
				\end{equation}
				for some $\vec{v} \in \setR^{n-1}$ and $\vec{u} \in \setR^{m-1}$.
				Then $\mat{Q} \in \mathcal{N}\bb{\mathscr{S}, 2}$.
			\end{lemma}

			We refer the reader to \remarkname~2 in~\cite{choudhary2014limitsBDambiguity} for correct parsing of the symbolic shorthand on the \rhs{} of \eqref{eqn:rank-2 nullspace}.
			We shall also need the following non-linear re-parameterization cum decomposition result (borrowed from \partname~I of this paper) to serve as a building block for constructing adversarial instances of input signals for which deconvolution fails the identifiability test.
			We point out the symbolic connection to the transformation in \eqref{eqn:rotational transform} and the representation in \eqref{eqn:rank-2 nullspace}.

			\begin{lemma}[from~\cite{choudhary2014limitsBDambiguity}]
				\label{lem:finite quotient set}
				Let $d \geq 2$ be an arbitrary integer and $\vec{w} \in \set{\vec{w}' \in \setR^{d}}{\vec{w}'(1) \neq 0, \vec{w}'(d) \neq 0}$ be an arbitrary vector.
				The quotient set $\mathcal{Q}_{\sim}\bb{\vec{w}, d}$ defined as
				\makeatletter
					\if@twocolumn
						\begin{equation}
							\begin{split}
								\mathcal{Q}_{\sim}\bb{\vec{w}, d}
								& \triangleq	\mleft\{\bb{\vec{w}_{\ast},\gamma} \in \setR^{d-1} \times \setA \, \middle| \vphantom{\begin{bmatrix} \cos \gamma \\ \sin \gamma \end{bmatrix}} \mright.	\\
								& \qquad \quad \mleft. \vec{w} =	\begin{bmatrix}
																		\vec{w}_{\ast}	&	0	\\
																		0	&	-\vec{w}_{\ast}
																	\end{bmatrix}
																	\begin{bmatrix}
																		\cos \gamma	\\
																		\sin \gamma
																	\end{bmatrix} \mright\},
							\end{split}
						\end{equation}
					\else
						\begin{equation}
							\mathcal{Q}_{\sim}\bb{\vec{w}, d}
							\triangleq	\set{\bb{\vec{w}_{\ast},\gamma} \in \setR^{d-1} \times \setA}
										{\vec{w} =	\begin{bmatrix}
														\vec{w}_{\ast}	&	0	\\
														0	&	-\vec{w}_{\ast}
													\end{bmatrix}
													\begin{bmatrix}
														\cos \gamma	\\
														\sin \gamma
													\end{bmatrix}},
						\end{equation}
					\fi
				\makeatother
				is finite (possibly empty) with cardinality at most $\bb{2d-2}$.
				If $d$ is an even integer then $\mathcal{Q}_{\sim}\bb{\vec{w}, d}$ is non-empty.
			\end{lemma}

			We briefly recap the pathological cases discussed in \partname~I, \sectionname~IV-A that we shall exclude from consideration in rest of the paper.
			This serves to remove the most straightforward delay ambiguities at an intuitive level, as well as to simplify the arguments in our proofs by excluding special cases that are best considered separately.
			It was shown in \partname~I that for identifiability of $\bb{\vec{x}_{\ast}, \vec{y}_{\ast}} \in \setR^{m} \times \setR^{n}$ within $\setR^{m} \times \setR^{n}$, it is necessary that both $\cc{\vec{x}_{\ast}(m), \vec{y}_{\ast}(1)} \neq \cc{0}$ and $\cc{\vec{x}_{\ast}(1), \vec{y}_{\ast}(n)} \neq \cc{0}$ must be true.
			In the sequel, we shall consider the stronger restriction $0 \not \in \cc{\vec{x}_{\ast}(1), \vec{x}_{\ast}(m), \vec{y}_{\ast}(1), \vec{y}_{\ast}(n)}$ to automatically eliminate the pathological cases.
			Implicitly, this is also the reason for requiring $\vec{w}(1) \neq 0$ and $\vec{w}(d) \neq 0$ for any vector $\vec{w} \in \setR^{d}$ in the premise of \lemmaname~\ref{lem:finite quotient set} as well as in the definition of the canonical-sparse domain $\Kzero\bb{\Lambda,d}$ in \eqref{eqn:sparse domain}.

		\subsection{Canonical-sparse Blind Deconvolution}
			\label{sec:sparse deconv}
			To be consistent with the notation for canonical-sparse cones in \eqref{eqn:sparse domain}, we denote the set of unconstrained non-pathological $d$ dimensional vectors by
			\begin{equation}
				\mathcal{K}\bb{\emptyset,d}	\triangleq \set{\vec{w} \in \setR^{d}}{\vec{w}\bb{1} \neq 0, \vec{w}\bb{d} \neq 0}.
			\end{equation}
			For subsequent comparison we recall below, the unidentifiability result from \partname~I of the paper adapted for the domain $\mathcal{K}\bb{\emptyset,d}$.

			\begin{theorem}[adapted from~\cite{choudhary2014limitsBDambiguity}]
				\label{thm:ae unident}
				Let $m,n \geq 4$ be even integers and $\mathcal{K} = \mathcal{K}\bb{\emptyset,m} \times \setR^{n}$.
				For any $\vec{x} \in \mathcal{K}\bb{\emptyset,m}$, $\bb{\vec{x}, \vec{y}} \in \mathcal{K}$ is unidentifiable almost everywhere \wrt~any measure over $\vec{y}$ that is absolutely continuous \wrt~the $n$ dimensional Lebesgue measure.
			\end{theorem}

			\begin{figure}
				\centering
				\includegraphics[width=0.8\figwidth]{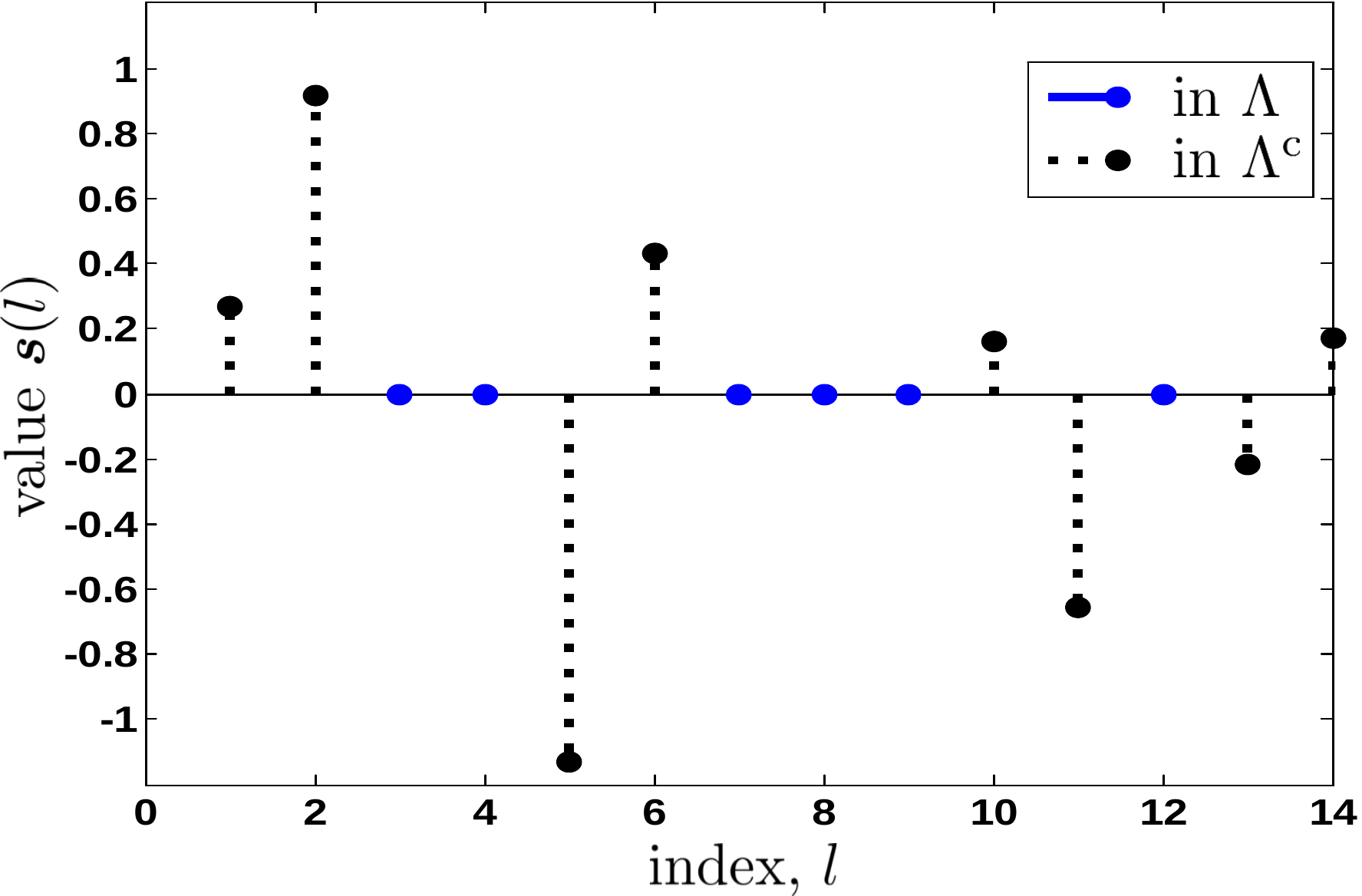}
				\caption{An arbitrary vector $\vec{s} \in \Kzero\bb{\Lambda, d}$ with $\Lambda = \cc{3,4,7,8,9,12}$ and $d = 14$.
				Every $\vec{s} \in \Kzero\bb{\Lambda, d}$ is zero on the index set $\Lambda$ (indicated by blue dots).
				Heights of the black dashed stems, indicating values on $\Lambda^{\comp}$, can vary across different vectors in $\Kzero\bb{\Lambda, d}$.}
				\label{fig:sample-K0}
			\end{figure}
			In the presence of a sparsity prior on $\vec{y} \in \setR^{n}$, \theoremname~\ref{thm:ae unident} does not apply anymore, since a sparsity prior is necessarily generated from a measure that is \emph{not} absolutely continuous \wrt~the $n$ dimensional Lebesgue measure.
			Assuming that the sparsity prior is \wrt~the canonical basis, we prove the following unidentifiability result (recall that $\mathcal{K}/\IdR$ denotes the set of all equivalence classes induced by the equivalence relation $\IdR\bb{\cdot,\cdot}$ on $\mathcal{K}$).

			\begin{theorem}
				\label{thm:sparse unident moderate}
				Let $m,n \geq 5$ be arbitrary integers.
				For any given index sets $\emptyset \neq \Lambda_{1} \subseteq \cc{3,4,\dotsc,m-2}$ and $\emptyset \neq \Lambda_{2} \subseteq \cc{3,4,\dotsc,n-2}$, let $\mathcal{K} = \Kzero\bb{\Lambda_{1},m} \times \Kzero\bb{\Lambda_{2},n}$ and define $p_{j} \triangleq \card{\Lambda_{j} \bigcup \bb{\Lambda_{j} - 1}}$ for $j \in \cc{1,2}$.
				Then there exists a set $\mathcal{G}_{\ast} \subseteq \mathcal{K}/\IdR$ of dimension $\bb{m+n-1 - p_{1} - p_{2}}$ such that every $\bb{\vec{x}, \vec{y}} \in \mathcal{G}_{\ast}$ is unidentifiable.
			\end{theorem}

			\begin{IEEEproof}
				\appendixname~\ref{sec:sparse unident moderate proof}.
			\end{IEEEproof}

			As a visualization aid, an arbitrary vector $\vec{s}$ in the canonical-sparse cone $\Kzero\bb{\Lambda,d} = \Kzero\bb{\cc{3,4,7,8,9,12}, 14}$ is shown in \figurename~\ref{fig:sample-K0}.
			A couple of comments about the premise of \theoremname~\ref{thm:sparse unident moderate} are in order.
			Let $\bb{\vec{x}_{\ast}, \vec{y}_{\ast}} \in \mathcal{K}$ denote an unidentifiable pair in the feasible set of \theoremname~\ref{thm:sparse unident moderate}.
			\begin{enumerate}
				\item	Our proof technique hinges on an adversarial construction of $\vec{x}_{\ast}\bb[\big]{\Lambda_{1} \bigcup \bb{\Lambda_{1} - 1}}$ and $\vec{y}_{\ast}\bb[\big]{\Lambda_{2} \bigcup \bb{\Lambda_{2} - 1}}$.
						By feasibility of $\bb{\vec{x}_{\ast}, \vec{y}_{\ast}}$, $\cc{\vec{x}_{\ast}\bb{1}, \vec{x}_{\ast}\bb{m}, \vec{y}_{\ast}\bb{1}, \vec{y}_{\ast}\bb{n}} \not \owns 0$ needs to be satisfied, which might conflict with our adversarial construction if $\cc{1,m} \bigcap \bb[\big]{\Lambda_{1} \bigcup \bb{\Lambda_{1} - 1}} \neq \emptyset$ or $\cc{1,n} \bigcap \bb[\big]{\Lambda_{2} \bigcup \bb{\Lambda_{2} - 1}} \neq \emptyset$ holds.
						Clearly, both of these scenarios are rendered impossible, if we insist on $\Lambda_{1} \subseteq \cc{3,4,\dotsc,m-2}$ and $\Lambda_{2} \subseteq \cc{3,4,\dotsc,n-2}$.
				\item	We insist on $\Lambda_{1} \neq \emptyset$ and $\Lambda_{2} \neq \emptyset$ to have a strictly non-trivial realization of the \emph{sparse} blind deconvolution problem, \ie~an instance which violates some assumption of \theoremname~\ref{thm:ae unident} other than model orders $m$ and $n$ being even.
						It is easy to see that if $\Lambda = \emptyset$ in \eqref{eqn:sparse domain} then $\Kzero\bb{\Lambda,d} = \mathcal{K}\bb{\emptyset,d}$ admits a non-zero $d$ dimensional Lebesgue measure.
						Hence, if either $\Lambda_{1}$ or $\Lambda_{2}$ is empty in \theoremname~\ref{thm:sparse unident moderate}, then the sparse blind deconvolution problem instance so generated will fall under the purview of \theoremname~\ref{thm:ae unident} (assuming even $m$ and $n$).
						Technically speaking, \theoremname~\ref{thm:sparse unident moderate} is valid even if $\Lambda_{1}$ or $\Lambda_{2}$ is empty (provided that $m,n \geq 4$), however its implications are weaker than that of \theoremname~\ref{thm:ae unident} when $m$ and $n$ are even.
						Furthermore, note that the requirements $\Lambda_{1} \neq \emptyset$ and $\Lambda_{2} \neq \emptyset$ respectively imply $m \geq 5$ and $n \geq 5$.
			\end{enumerate}

			We note that the assumptions of \theoremname~\ref{thm:sparse unident moderate} imply $p_{1} \leq \bb{m-3}$ and $p_{2} \leq \bb{n-3}$ so that the unidentifiable subset of $\mathcal{K}/\IdR$ is at least 5 dimensional.
			In particular, the unidentifiable set in \theoremname~\ref{thm:sparse unident moderate} is always non-trivial.
			We also note that the canonical-sparse feasible domain of \theoremname~\ref{thm:sparse unident moderate} is far more structured than the feasible set of \theoremname~\ref{thm:ae unident} and hence, the set of all unidentifiable inputs in \theoremname~\ref{thm:sparse unident moderate} is much smaller.
			Nonetheless, the canonical-sparsity structure is not strong enough to guarantee identifiability for all canonical-sparse vectors.
			Additionally, note that $\Lambda_{1}$ and $\Lambda_{2}$ denote \emph{sets of zero indices}, so that larger cardinality of $\Lambda_{1}$ or $\Lambda_{2}$ implies a sparser problem instance.
			Furthermore, we note that \corollaryname~\ref{cor:mixed unident moderate} straightforwardly follows from \theoremname~\ref{thm:sparse unident moderate} by setting $\Lambda_{1} = \Lambda$ and $\Lambda_{2} = \emptyset$.
			However, the result in \corollaryname~\ref{cor:mixed unident moderate} is not tight for $n \geq 4$ being an even integer.
			We remedy this shortcoming in \sectionname~\ref{sec:mixed extensions} by directly specializing \theoremname~\ref{thm:ae unident} to $\mathcal{K} = \Kzero\bb{\Lambda,m} \times \setR^{n}$ (the feasible set of \corollaryname~\ref{cor:mixed unident moderate}) to yield \corollaryname~\ref{cor:mixed unident everywhere}.

			Exploiting the equivalence between bilinear inverse problems and rank one matrix recovery problems~\cite{choudhary2013bilinear} (in the current context, equivalence between \problemsname~\eqref{prob:find_xy} and \eqref{prob:rank}) the result of \theoremname~\ref{thm:sparse unident moderate} can be interpreted as evidence of the null space of the convolution operator admitting a large number of simultaneously canonical-sparse and low-rank matrices.
			For rank one \emph{matrix completion} problems~\cite{candes2009exact}, it is relatively straightforward to see that a random sampling operator on a sparse rank one matrix will return zeros on most samples, thus rendering it impossible to distinguish the rank one matrix in question from the all zero matrix.
			However, the same observation is not at all straightforward for a rank one \emph{matrix recovery} problem~\cite{gross2011recovering} when the sampling operator is fixed to the lifted linear convolution operator.
			\theoremname~\ref{thm:sparse unident moderate} asserts that this is indeed true and the lifted linear convolution operator $\mathscr{S}\bb{\cdot}$ admits a large number of non-zero canonical-sparse matrices within its rank two null space.

			It is important to note that \theoremsname~\ref{thm:ae unident} and~\ref{thm:sparse unident moderate} are of different flavors and are not comparable since they make different assumptions on the feasible domain.
			In particular, neither theorem universally implies the other even if we consider special cases for each of them.
			To make this point more explicit, we make the following observations.
			\begin{enumerate}
				\item	\theoremname~\ref{thm:ae unident} asserts almost everywhere unidentifiability within the feasible set $\mathcal{K}\bb{\emptyset,m} \times \setR^{n}$, but this does not imply the conclusions of \theoremname~\ref{thm:sparse unident moderate}, since the feasible set in the latter theorem is $\Kzero\bb{\Lambda_{1},m} \times \Kzero\bb{\Lambda_{2},n}$, which is a measure zero set \wrt~the $\bb{m+n}$ dimensional Lebesgue measure associated with the Cartesian product space $\mathcal{K}\bb{\emptyset,m} \times \setR^{n}$.
				\item	In \theoremname~\ref{thm:sparse unident moderate}, even if we assume the model orders $m$ and $n$ to be even and the zero index sets $\Lambda_{1}$ and $\Lambda_{2}$ to be empty (so that the feasible set $\Kzero\bb{\Lambda_{1},m} \times \Kzero\bb{\Lambda_{2},n}$ is equal to $\mathcal{K}\bb{\emptyset,m} \times \setR^{n}$ almost everywhere \wrt~the $\bb{m+n}$ dimensional Lebesgue measure), we can only draw the conclusion that there exists a $\bb{m+n}$ dimensional unidentifiable subset (call it $\mathcal{G}_{\text{unID}}$) of the feasible set.
				This is clearly insufficient to show almost everywhere unidentifiability (as claimed by \theoremname~\ref{thm:ae unident}) since the complement of $\mathcal{G}_{\text{unID}}$ could be a $\bb{m+n}$ dimensional set as well, admitting non-zero $\bb{m+n}$ dimensional Lebesgue measure.
				\item	If $\Lambda_{1} = \Lambda_{2} = \emptyset$ is considered with even model orders $m,n \geq 4$ so that the feasible sets in \theoremsname~\ref{thm:ae unident} and~\ref{thm:sparse unident moderate} are equal almost everywhere \wrt~the $\bb{m+n}$ dimensional Lebesgue measure, then the conclusion of \theoremname~\ref{thm:ae unident} is stronger, since almost everywhere unidentifiability over $\mathcal{K}\bb{\emptyset,m} \times \setR^{n}$ automatically implies the existence of a $\bb{m+n}$ dimensional unidentifiable subset.
			\end{enumerate}

		\subsection{A Mixed Extension}
			\label{sec:mixed extensions}
			It is possible to fuse the ideas from \theoremsname~\ref{thm:ae unident} and~\ref{thm:sparse unident moderate} to strengthen the conclusion of \corollaryname~\ref{cor:mixed unident moderate} on the domain $\mathcal{K} = \Kzero\bb{\Lambda,m} \times \setR^{n}$, \ie~on a domain formed by the Cartesian product of a canonical-sparse and a non-sparse cone.

			\begin{corollary}
				\label{cor:mixed unident everywhere}
				Let $m \geq 5$ be an arbitrary integer and $n \geq 4$ be an even integer.
				For any given index set $\emptyset \neq \Lambda \subseteq \cc{3,4,\dotsc,m-2}$, let $\mathcal{K} = \Kzero\bb{\Lambda,m} \times \setR^{n}$ and define $p \triangleq \card{\Lambda \bigcup \bb{\Lambda - 1}}$.
				Then there exists a set $\mathcal{H}' \subseteq \Kzero\bb{\Lambda,m}$ of dimension $\bb{m-p}$ such that $\forall \vec{x} \in \mathcal{G}'$, $\bb{\vec{x}, \vec{y}} \in \mathcal{K}$ is unidentifiable almost everywhere \wrt~any measure over $\vec{y}$ that is absolutely continuous \wrt~the $n$ dimensional Lebesgue measure.
			\end{corollary}

			\begin{IEEEproof}
				\appendixname~\ref{sec:mixed unident everywhere proof}.
			\end{IEEEproof}

			Note that \corollaryname~\ref{cor:mixed unident everywhere} is a non-trivial extension of \theoremname~\ref{thm:ae unident}, since \definitionname~\ref{defn:identifiability} defines identifiability of an input signal pair \emph{within some feasible set} $\mathcal{K}$ and this set is different for \theoremname~\ref{thm:ae unident} and \corollaryname~\ref{cor:mixed unident everywhere}.
			In particular, if $\bb{\vec{x}_{\ast}, \vec{y}_{\ast}} \in \mathcal{K}$ is an unidentifiable pair in the feasible set of \corollaryname~\ref{cor:mixed unident everywhere}, then the proof of the result involves the construction of a candidate adversarial input $\vec{x}_{\ast} \in \Kzero\bb{\Lambda,m}$ as in \theoremname~\ref{thm:sparse unident moderate} as well as the vector $\vec{y}_{\ast} \in \setR^{n}$ as in \theoremname~\ref{thm:ae unident}.

			Through the unidentifiability results in this section, we have attempted to \emph{quantify} the ill-posedness of blind deconvolution under canonical-sparsity constraints, and illustrate the geometric reasons behind this unfavorable sparsity-ambiguity trade-off behavior.
			The rule of thumb for determining the sparsity-ambiguity trade-off for canonical-sparse blind deconvolution is well illustrated by comparing the conclusions of \theoremname~\ref{thm:ae unident} and \corollaryname~\ref{cor:mixed unident everywhere}.
			Roughly speaking, if the set of zero indices for the signal vector $\vec{x} \in \setR^{m}$ has cardinality $\card{\Lambda}$ then the dimension of the set of unidentifiable choices for $\vec{x}$ is reduced from $m$ in \theoremname~\ref{thm:ae unident} to somewhere between $\bb{m-\card{\Lambda}-1}$ and $\bb{m - 2\card{\Lambda}}$ in \corollaryname~\ref{cor:mixed unident everywhere}.
			The main message is that if an application exhibits a bilinear observation model of linear convolution and the additional application specific structure is that of canonical sparsity of the unknown variables, then it is necessary to drastically revise the system design specifications for any hope of signal identifiability.
			Such revision may incorporate some form of randomized precoding of the unknowns, so that the \emph{effective} bilinear operator governing the observation model looks substantially different from the convolution operator (\eg~the Gaussian random precoding used in~\cite{ahmed2012blind}).
			Alternatively, sparsity in non-canonical bases could also be helpful (\eg~Rademacher random vector signal model used in~\cite{choudhary2013bilinear} and the constant modulus based problem instances in \cite{johnson1998blind}).

	\section{Unidentifiability Results with Coding}
		\label{sec:results with coding}
		Structural information about unknown input signals is almost exclusively a result of specific application dependant data or system deployment architectures.
		In this section, we consider an abstraction for a problem in multi-hop channel estimation~\cite{choudhary2012sparse,richard2008sparse} and analyze other subspace based structural priors relevant to this application.
		Studying constrained adversarial realizations of these abstractions has utility far beyond communication systems.
		For example, when communication subsystems are parts of a larger complex system involving sensing and navigation for autonomous vehicles~\cite{hollinger2011communicationprotocolsunderwater,choudhary2014activetargetdetection,hollinger2011autonomousdatacollection,freitag2001integratedacousticcommunication,choudhary2014ATDwithNav} then non-linear inter-system interactions may become unavoidable.
		In this section, we extend our unidentifiability results from \sectionname~\ref{sec:results without coding} to simple forms of coding across the transmitters for the second-hop channel estimation problem.
		\sectionname~\ref{sec:simple coding} considers the somewhat idealized, but simpler to interpret, scenario of repetition coding across transmitters and states unidentifiability results for both unstructured as well as canonical-sparse channels.
		Rather than explicitly proving the results in \sectionname~\ref{sec:simple coding}, we note that they are important special cases of the more general unidentifiability result in \sectionname~\ref{sec:partial cooperation} for certain families of partially cooperative codes (\theoremname~\ref{thm:sparse cooperative coding unident}) that we prove in \appendixname~\ref{sec:sparse cooperative coding unident proof}.
		Finally, \corollaryname~\ref{cor:sparse geometric unident} states yet another important special case of \theoremname~\ref{thm:sparse cooperative coding unident} pertaining to geometrically decaying subspace sparse signals, previously considered in~\cite{choudhary2014subspaceblind}.
		In the spirit of using commonly employed notation, we shall denote the unknown pair $\bb{\vec{x}, \vec{y}}$ by the pair $\bb{\vec{g}, \vec{h}}$ with $\vec{g}$ pertaining to a topological configuration of the network and $\vec{h}$ representing the channel impulse response as described next.

		\begin{figure}
			\centering
			\includegraphics[width=0.8\figwidth]{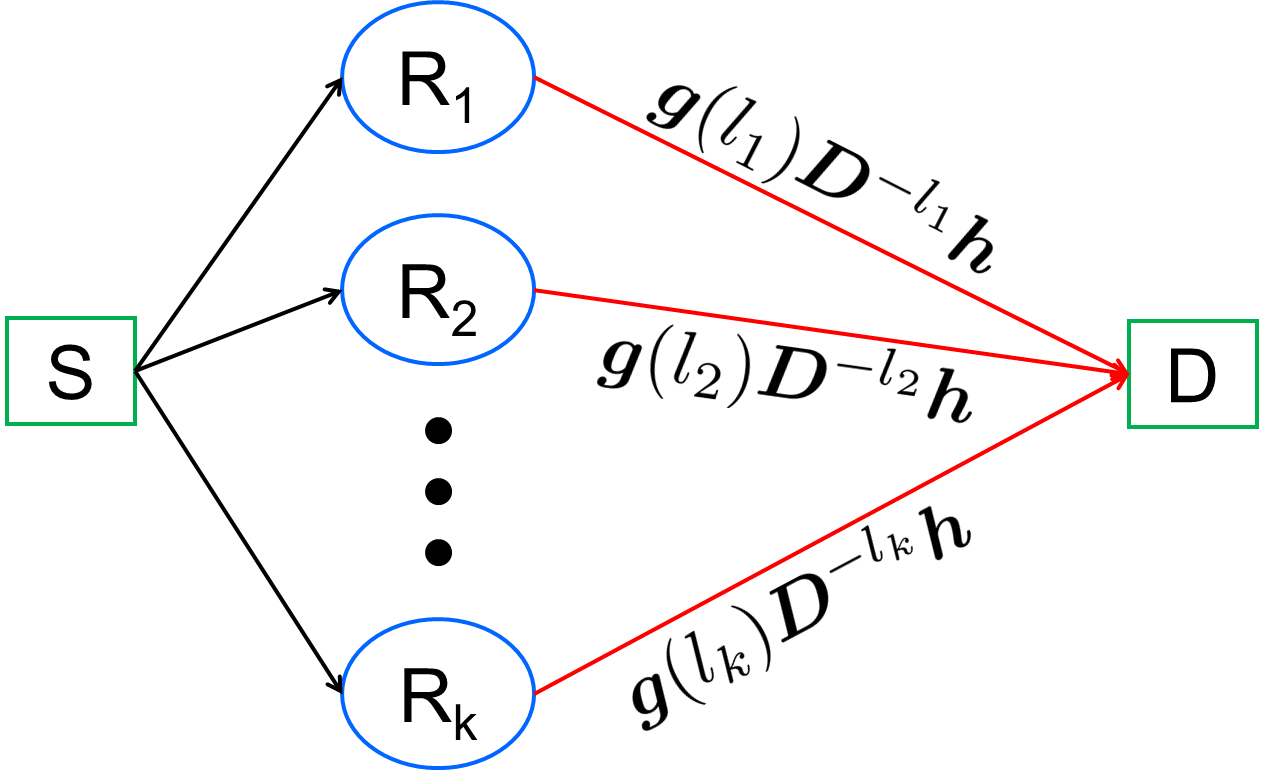}
			\caption{Two hop relay assisted communication link topology between the source S and the destination D with $k$ intermediate relays R\textsubscript{1}, R\textsubscript{2}, \dots, R\textsubscript{k}.
			The $j^{\thp}$ SISO channel is scaled by $\vec{g}\bb{l_{j}}$ and delayed by $l_{j}$ units (represented as $\vec{g}\bb{l_{j}} \mat{D}^{-l_{j}} \vec{h}$) for $1 \leq j \leq k$.}
			\label{fig:topology}
		\end{figure}
		In a previous paper~\cite{richard2008sparse}, we developed a structured channel model based on the \emph{multichannel approximation} for the second hop in a relay assisted communication topology shown in \figurename~\ref{fig:topology}.
		Specifically, we consider $k$ relays R\textsubscript{1}, R\textsubscript{2}, \dots, R\textsubscript{$k$} and let $\vec{h} \in \setR^{n}$ denote the \emph{propagation delay and power adjusted} SISO channel impulse response (CIR) common to each relay-destination R\textsubscript{$j$} $\rightarrow$ D channel, $j \in \cc{1,2,\dots,k}$.
		Thus, the effective CIR at the destination is formed by a superposition of $k$ distinct delayed scalar multiples of the vector $\vec{h}$.
		If the maximum propagation delay of R\textsubscript{$k$} relative to R\textsubscript{1} is upper bounded by $m$, then there exists a vector $\vec{g} \in \setR^{m}$ and an index subset $\cc{l_{1}, l_{2}, \dotsc, l_{k}} \subseteq \cc{1,2,\dotsc,m}$ (with $l_{1} = 1$) such that the propagation delay adjusted SISO CIR for the R\textsubscript{$j$} $\rightarrow$ D channel is $\vec{g}\bb{l_{j}} \vec{h}$ and that $\vec{g}\bb{l} = 0$, $\forall l \not \in \cc{l_{1}, l_{2}, \dotsc, l_{k}}$.
		With these definitions and letting $\mat{D}^{-l} \in \cc{0,1}^{\bb{m+n-1} \times n}$ denote the Toeplitz matrix representation for delay by $l$ units $\forall 1 \leq l \leq m$, the effective CIR at the destination is given by the linear convolution $\vec{z} = \sum_{j=1}^{k} \vec{g}\bb{l_{j}} \mat{D}^{-l_{j}} \vec{h} = \vec{g} \star \vec{h} \in \setR^{m+n-1}$ and the channel estimation problem at the destination is to recover the vector pair $\bb{\vec{g}, \vec{h}}$ from the observation of $\vec{z}$.
		Since the vectors $\vec{g}$ and $\vec{h}$ are not arbitrary, but have physical interpretations leading to structural restrictions, we can incorporate this knowledge by a constraint of the form of $\bb{\vec{g}, \vec{h}} \in \mathcal{K}$ for some set $\mathcal{K} \subseteq \setR^{m} \times \setR^{n}$.
		For example, if $\vec{h}$ represents a SISO CIR for underwater acoustic communication~\cite{berger2010sparse,li2007estimation} (or, more generally, for wide-band communication~\cite{molisch2009UltraWideBand}), then $\vec{h}$ has been observed to exhibit canonical-sparsity.
		Now, associating the pair $\bb{\vec{g}, \vec{h}}$ with a feasible solution to \problemname~\eqref{prob:find_xy}, the success of the channel estimation problem at the destination is critically contingent upon the identifiability of $\bb{\vec{g}, \vec{h}}$ as a solution to \problemname~\eqref{prob:find_xy} by the criterion in \definitionname~\ref{defn:identifiability}.

		As with our prior results, herein we shall consider $\mathcal{K} = \mathcal{D}_{1} \times \mathcal{D}_{2}$ to be a separable cone in $\setR^{m} \times \setR^{n}$ with $\vec{g} \in \mathcal{D}_{1} \subseteq \setR^{m}$ and $\vec{h} \in \mathcal{D}_{2} \subseteq \setR^{n}$.
		In many applications, it is typical to have $\mathcal{D}_{1}$ (respectively $\mathcal{D}_{2}$) to be a low-dimensional subset of $\setR^{m}$ (respectively $\setR^{n}$).
		An important interpretation of $\vec{g} \in \mathcal{D}_{1}$ is that $\vec{g}$ can be considered as a coded representation (of ambient dimension $m$) of the unknown whose actual dimension is much smaller (equal to that of the Hausdorff dimension of $\mathcal{D}_{1}$).
		However, this coded representation is invoked to help strengthen the identifiability properties of the problem, rather than to help with error correction in the presence of observation noise.
		This is a somewhat different interpretation of coding than the classical goal of correcting noise induced errors.
		Operationally, both interpretations rely on redundancy albeit for different purposes.
		
		\subsection{Repetition Coding}
			\label{sec:simple coding}
			Suppose that a \emph{few} of the relays in \figurename~\ref{fig:topology} cooperatively decide to maintain the same relative transmission powers and phases and this decision is communicated to the destination for use as side information for channel estimation.
			A physical motivation for allowing \emph{few} rather than \emph{all} relays to cooperate, may be attributed to infeasibility of such coordination for widely separated relays and/or for a large number of relays.
			Mathematically, this translates to the use of a smaller feasible set $\mathcal{K}$ by virtue of a stricter structural restriction on $\vec{g}$ via the set $\mathcal{D}_{1}$ that is smaller than $\setR^{m}$.
			Specifically, let $\Lambda' \subseteq \cc{l_{1}, l_{2}, \dotsc, l_{k}} \subseteq \cc{1,2,\dotsc,m}$ be the index set of propagation delays corresponding to the cooperating relays.
			Assuming that the only structure on $\vec{g} \in \mathcal{D}_{1}$ is that induced by the cooperating relays, we define the repetition coded domain as $\mathcal{D}_{1} = \Kone\bb{\Lambda', m}$ where $\bb{\Lambda', m}$ act as parameters to the following parametrized definition of a family of \emph{cones} for any dimension $d \geq 2$ and any index set $\Lambda \subseteq \cc{1,2,\dotsc,d}$,
			\makeatletter
				\if@twocolumn
					\begin{equation}
						\begin{split}
							\Kone\bb{\Lambda, d}
							& \triangleq	\mleft\{\vec{w} \in \setR^{d} \, \middle| \, \vec{w}\bb{1} \neq 0, \vec{w}\bb{d} \neq 0, \mright.	\\
							& \qquad \quad \mleft. \exists c \in \setR \setminus \cc{0}, \text{ such that } \vec{w}\bb{\Lambda} = c\vec{1} \mright\}.
						\end{split}
						\label{eqn:rep code defn}
					\end{equation}
				\else
					\begin{equation}
						\Kone\bb{\Lambda, d}	 \triangleq \set{\vec{w} \in \setR^{d}}{\vec{w}\bb{1} \neq 0, \vec{w}\bb{d} \neq 0, \exists c \in \setR \setminus \cc{0}, \text{ such that } \vec{w}\bb{\Lambda} = c\vec{1}}.
						\label{eqn:rep code defn}
					\end{equation}
				\fi
			\makeatother
			In other words, the set $\mathcal{D}_{1}$ imposes the structure that each of its member vectors has the same non-zero value $c$ on the index subset $\Lambda'$; the value of $c$ being allowed to \emph{vary across members}.
			Even for a large number of cooperating relays (as measured by the cardinality $\card{\Lambda'}$), \corollariesname~\ref{cor:repetition coding unident} and~\ref{cor:sparse repetition coding unident} stated below imply unidentifiability results for $\vec{h} \in \setR^{n}$ (unstructured channel) and $\vec{h} \in \Kzero\bb{\Lambda'', n}$ (canonical-sparse channel), respectively.
			\corollaryname~\ref{cor:repetition coding unident} makes a statement for the repetition coded domain $\Kone\bb{\Lambda',m}$ that is analogous to the conclusions drawn by \corollaryname~\ref{cor:mixed unident moderate} for the canonical-sparse domain $\Kzero\bb{\Lambda,m}$.
			The main technical difference between \corollariesname~\ref{cor:repetition coding unident} and~\ref{cor:mixed unident moderate} is that they require different constructions for a candidate unidentifiable input signal since the feasible domain $\mathcal{K}$ is different across these results.
			We defer further technical details to the proof of \theoremname~\ref{thm:sparse cooperative coding unident} in \appendixname~\ref{sec:sparse cooperative coding unident proof}.
			\corollaryname~\ref{cor:sparse repetition coding unident} can be interpreted as extending the unidentifiability results of \theoremname~\ref{thm:sparse unident moderate} from the canonical-sparse product domain $\Kzero\bb{\Lambda_{1},m} \times \Kzero\bb{\Lambda_{2},n}$ to the mixed product of repetition coded and canonical-sparse domains $\Kone\bb{\Lambda',m} \times \Kzero\bb{\Lambda'',n}$.
			Roughly speaking, because we only consider \emph{separable} domains $\mathcal{K}$, one can simply fuse the different constructions of a candidate adversarial signal $\vec{x}_{\ast} \in \Kone\bb{\Lambda',m}$ from \corollaryname~\ref{cor:repetition coding unident} and $\vec{y}_{\ast} \in \Kzero\bb{\Lambda'',n}$ from \corollaryname~\ref{cor:mixed unident moderate} to produce a candidate unidentifiable pair $\bb{\vec{x}_{\ast}, \vec{y}_{\ast}} \in \Kone\bb{\Lambda',m} \times \Kzero\bb{\Lambda'',n}$ for \corollaryname~\ref{cor:sparse repetition coding unident}.
			This ability to fuse individual adversarial signal constructions on sets $\mathcal{D}_{1}$ and $\mathcal{D}_{2}$ to produce candidate unidentifiable signal pairs within the Cartesian product domain $\mathcal{K} = \mathcal{D}_{1} \times \mathcal{D}_{2}$ underlies the simplicity of arguments in the proof of \theoremname~\ref{thm:sparse cooperative coding unident}.
			At present, we cannot foresee any adaptations to the constructions used in the proof of \theoremname~\ref{thm:sparse cooperative coding unident} that would allow us to make analogous statements for non-separable domains $\mathcal{K}$ in a generalizable way.

			\begin{corollary}
				\label{cor:repetition coding unident}
				Let $m \geq 3$ and $n \geq 2$ be arbitrary integers.
				For any given index set $\emptyset \neq \Lambda' \subseteq \cc{2,3,\dotsc,m-1}$, let $\mathcal{K} = \Kone\bb{\Lambda',m} \times \setR^{n}$ and define $p \triangleq \card{\Lambda' \bigcup \bb{\Lambda' - 1}}$.
				Then there exists a set $\mathcal{G}_{\ast} \subseteq \mathcal{K}/\IdR$ such that every $\bb{\vec{g}, \vec{h}} \in \mathcal{K}$ is unidentifiable.
				If $\Lambda' \bigcap \bb{\Lambda' - 1} \neq \emptyset$ then $\mathcal{G}_{\ast}$ is of dimension at least $\bb{m+n-p}$, otherwise $\mathcal{G}_{\ast}$ is of dimension at least $\bb{m+n-p+1}$.
			\end{corollary}

			\begin{corollary}
				\label{cor:sparse repetition coding unident}
				Let $m \geq 3$ and $n \geq 5$ be arbitrary integers.
				For any given index sets $\emptyset \neq \Lambda' \subseteq \cc{2,3,\dotsc,m-1}$ and $\emptyset \neq \Lambda'' \subseteq \cc{3,4,\dotsc,n-2}$, let $\mathcal{K} = \Kone\bb{\Lambda',m} \times \Kzero\bb{\Lambda'',n}$ and let $p_{1} \triangleq \card{\Lambda' \bigcup \bb{\Lambda' - 1}}$ and $p_{2} \triangleq \card{\Lambda'' \bigcup \bb{\Lambda'' - 1}}$.
				Then there exists a set $\mathcal{G}_{\ast} \subseteq \mathcal{K}/\IdR$ such that every $\bb{\vec{g}, \vec{h}} \in \mathcal{G}_{\ast}$ is unidentifiable.
				If $\Lambda' \bigcap \bb{\Lambda' - 1} \neq \emptyset$ then $\mathcal{G}_{\ast}$ is of dimension at least $\bb{m+n - p_{1} - p_{2}}$, otherwise $\mathcal{G}_{\ast}$ is of dimension at least $\bb{m+n+1 - p_{1} - p_{2}}$.
			\end{corollary}

			We note that in contrast to the results for the canonical-sparse domain $\Kzero\bb{\Lambda,m}$ in \corollaryname~\ref{cor:mixed unident moderate}, the dimension of unidentifiable subset for the repetition coded domain $\Kone\bb{\Lambda',m}$ as given by \corollaryname~\ref{cor:repetition coding unident} depends on whether $\Lambda' \bigcap \bb{\Lambda' - 1}$ is non-empty, \ie~if atleast one pair of adjacent indices for $\vec{g} \in \Kone\bb{\Lambda',m}$ are repetition coded and hence belong to $\Lambda'$.
			\corollaryname~\ref{cor:sparse repetition coding unident} essentially says that even with a canonical-sparse prior structure on the channel vector $\vec{h}$ and nearly full cooperation between intermediate relays ($\card{\Lambda'} = \Theta\bb{m}$) via repetition coding on $\vec{g}$, there may be non-zero dimensional unidentifiable signal subsets in the feasible domain $\mathcal{K}$.
			We recall that the index subsets $\Lambda'$ and $\Lambda''$ in \corollaryname~\ref{cor:sparse repetition coding unident} are \textit{known} at the receiver and the unidentifiable signal set exists despite the availability of this side information.
			Thus, the repetition coding based system architecture laid out in this section is bound to fail in practice although it may be an intuitive choice.
			An important implication of \corollaryname~\ref{cor:sparse repetition coding unident} is that it is necessary to have $\Lambda' \bigcap \cc{1,m} \neq \emptyset$, which is equivalent to the requirement that either the first relay or the last relay (when ordered by increasing propagation delays) must be present in the cooperating subset of relays.

		\subsection{Partially Cooperative Coding}
			\label{sec:partial cooperation}
			The results in \sectionname~\ref{sec:simple coding} can be generalized to other codes.
			Consider the system described in \sectionname~\ref{sec:simple coding}, except that the cooperating relays decide on a different structural restriction for $\vec{g} \in \mathcal{D}_{1} \subseteq \setR^{m}$.
			As before, let $\Lambda' \subseteq \cc{1,2,\dots,m}$ denote the index set of propagation delays corresponding to the cooperating relays.
			We let $\mathcal{D}_{1} = \Kb\bb{\Lambda', m}$ denote a partially cooperative coded domain where the vector $\vec{b} \in \setR^{\card{\Lambda'}}$ denotes the cooperative code (alternatively, $\vec{b}$ could also be interpreted as a power-delay-profile across cooperating relays) and $\bb{\Lambda', m}$ are specific parameters to the following parametrized definition of a family of cones for any dimension $d \geq 2$ and any index set $\Lambda \subseteq \cc{1,2,\dotsc,d}$,
			\makeatletter
				\if@twocolumn
					\begin{equation}
						\begin{split}
							\Kb\bb{\Lambda, d}
							& \triangleq	\mleft\{\vec{w} \in \setR^{d} \, \middle| \, \vec{w}\bb{1} \neq 0, \vec{w}\bb{d} \neq 0, \mright.	\\
							& \qquad \quad \mleft. \exists c \in \setR \setminus \cc{0}, \text{ such that } \vec{w}\bb{\Lambda} = c\vec{b} \mright\}.
						\end{split}
						\label{eqn:cooperative code defn}
					\end{equation}
				\else
					\begin{equation}
						\Kb\bb{\Lambda, d}	 \triangleq \set{\vec{w} \in \setR^{d}}{\vec{w}\bb{1} \neq 0, \vec{w}\bb{d} \neq 0, \exists c \in \setR \setminus \cc{0}, \text{ such that } \vec{w}\bb{\Lambda} = c\vec{b}}.
						\label{eqn:cooperative code defn}
					\end{equation}
				\fi
			\makeatother
			As a visual example, \figurename~\ref{fig:sample-domain-type-1} shows an arbitrary vector $\vec{s}$ in the partially cooperative coded domain $\Kb\bb{\Lambda,d} = \Kb\bb{\cc{3,4,7,8,9,12}, 14}$ for the code vector $\vec{b} = \tpose{\bb{0.5, 0.835, -0.3, -0.5, -0.835, -0.15}}$.
			It is important to note that the code vector $\vec{b} \in \setR^{\card{\Lambda'}}$ is \emph{known} at the destination so that the decoder has explicit knowledge of the set $\mathcal{D}_{1} = \Kb\bb{\Lambda', m}$.
			Furthermore, the partially cooperative coded domain $\Kb\bb{\Lambda, d}$ unifies and generalizes both the repetition coded domain $\Kone\bb{\Lambda, d}$ and the canonical-sparse domain $\Kzero\bb{\Lambda, d}$.
			To see this, we first note that $\vec{b} = \vec{0}$ implies $\Kb\bb{\Lambda, d} = \Kzero\bb{\Lambda, d}$ by definition, and that the definitions \eqref{eqn:rep code defn} and \eqref{eqn:cooperative code defn} are equivalent for $\vec{b} = \vec{1}$.
			Secondly, if the code vector $\vec{b} \in \setR^{\card{\Lambda_{1} \bigcup \Lambda_{2}}}$ admits a partition of the form $\BB{\tpose{\vec{b}\bb{\Lambda_{1}}}, \tpose{\vec{b}\bb{\Lambda_{2}}}} = \BB{\tpose{\vec{0}}, \tpose{\vec{1}}}$ for some \emph{disjoint} index subsets $\Lambda_{1}, \Lambda_{2} \subseteq \cc{1,2,\dots,d}$, then the definitions in \eqref{eqn:cooperative code defn}, \eqref{eqn:rep code defn} and \eqref{eqn:sparse domain} would imply that $\Kb\bb{\Lambda_{1} \bigcup \Lambda_{2}, d} = \Kzero\bb{\Lambda_{1}, d} \bigcap \Kone\bb{\Lambda_{2}, d}$.
			This can be checked directly by substitution.
			\makeatletter
				\if@twocolumn
					\begin{figure*}
						\centering
						\subfloat{\includegraphics[width=0.8\figwidth]{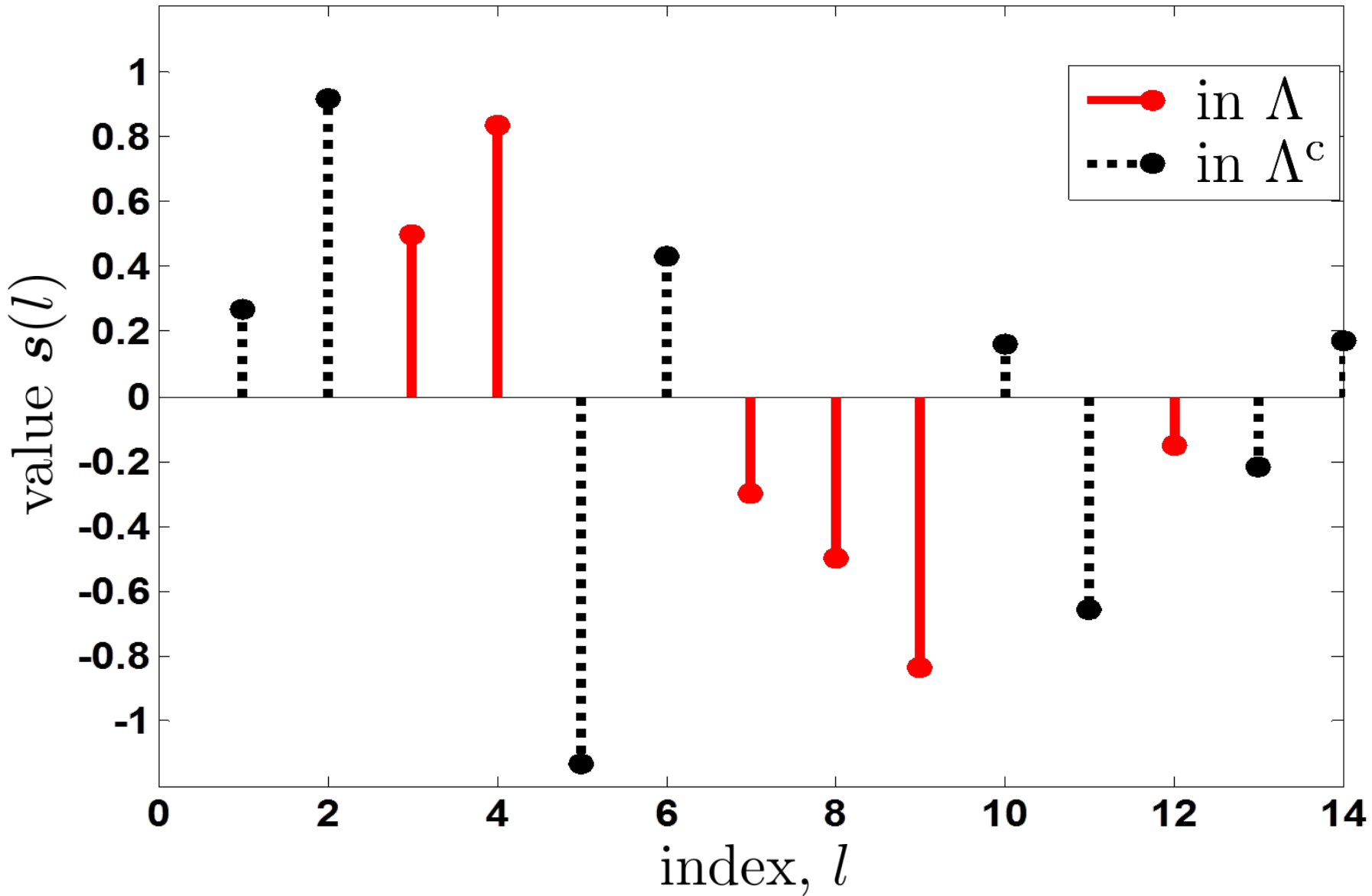}	\label{fig:sample-Kb}}
						\hspace{0.1\linewidth}
						\subfloat{\includegraphics[width=0.8\figwidth]{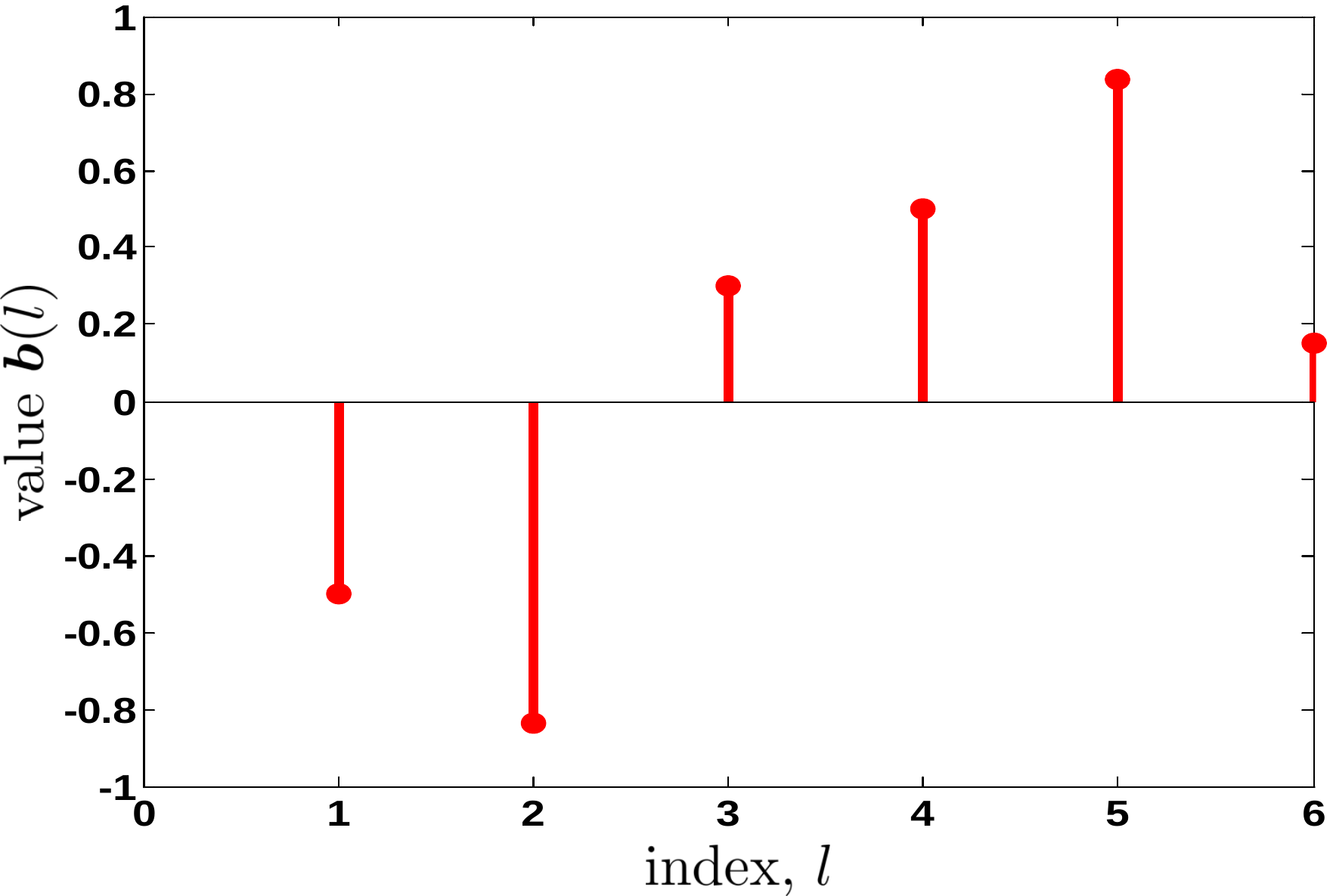}	\label{fig:example-b}}
						\caption{An arbitrary vector $\vec{s} \in \Kb\bb{\Lambda, d}$ with $\Lambda = \cc{3,4,7,8,9,12}$, $d = 14$ and $\vec{b} = \tpose{\bb{0.5, 0.835, -0.3, -0.5, -0.835, -0.15}}$.
						The specific vector $\vec{s}$ in the left plot satisfies $\vec{s}\bb{\Lambda} = c\vec{b}$ with $c = -1$, verifying \eqref{eqn:cooperative code defn} with code vector $\vec{b}$ as in the right plot.
						Every $\vec{s} \in \Kb\bb{\Lambda, d}$ is collinear with $\vec{b}$ on the index set $\Lambda$ (indicated by red stems) with $\Lambda$ representing the identities of the cooperating relays.
						Heights of the black dashed stems, indicating values on $\Lambda^{\comp}$, represent contributions from non-cooperating relays and can vary across different vectors in $\Kb\bb{\Lambda, d}$.}
						\label{fig:sample-domain-type-1}
					\end{figure*}
				\else
					\begin{figure}
						\centering
						\subfloat{\includegraphics[width=0.8\figwidth]{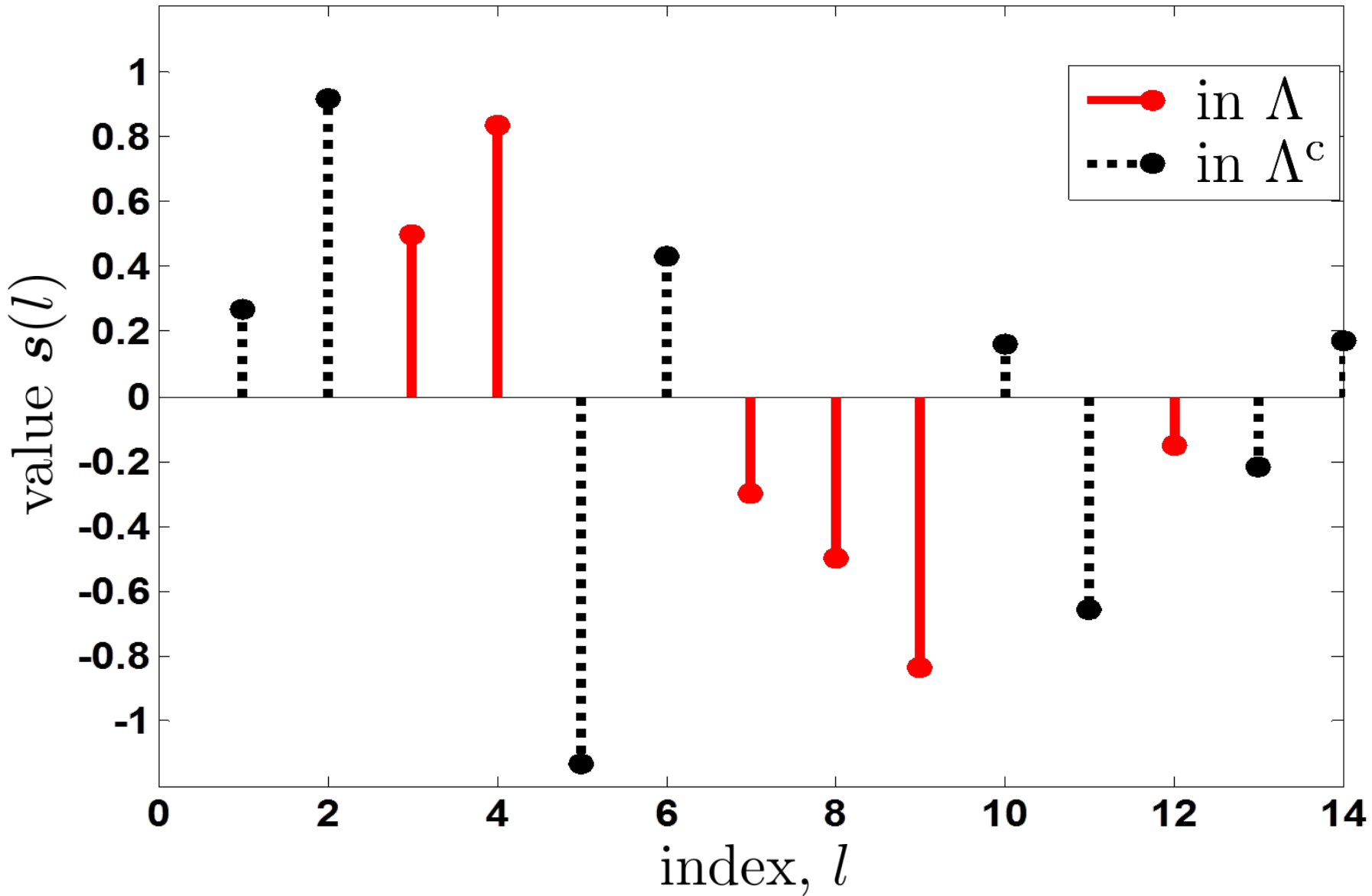}	\label{fig:sample-Kb}}
						\hspace{0.1\linewidth}
						\subfloat{\includegraphics[width=0.8\figwidth]{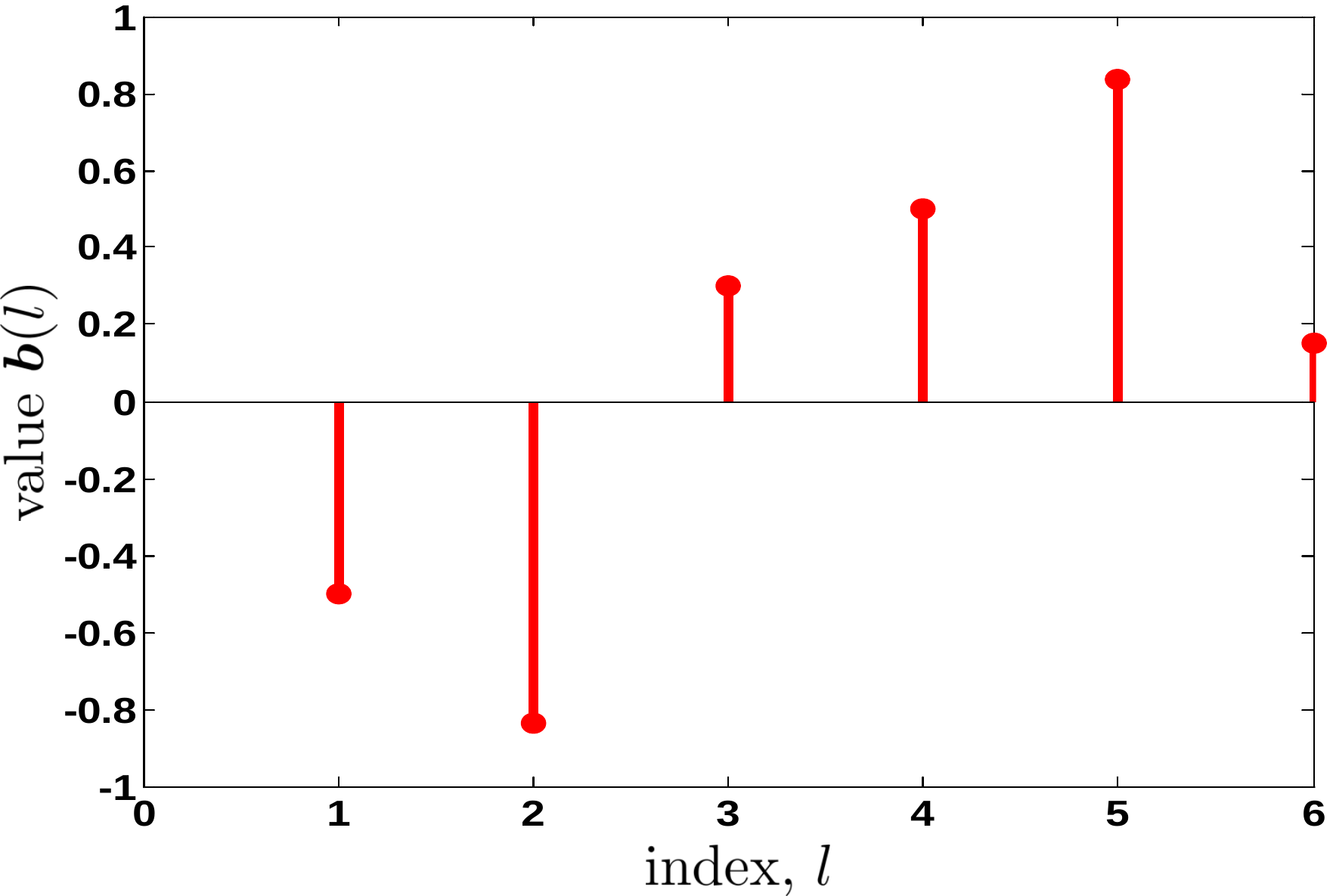}	\label{fig:example-b}}
						\caption{An arbitrary vector $\vec{s} \in \Kb\bb{\Lambda, d}$ with $\Lambda = \cc{3,4,7,8,9,12}$, $d = 14$ and $\vec{b} = \tpose{\bb{0.5, 0.835, -0.3, -0.5, -0.835, -0.15}}$.
						The specific vector $\vec{s}$ in the left plot satisfies $\vec{s}\bb{\Lambda} = c\vec{b}$ with $c = -1$, verifying \eqref{eqn:cooperative code defn} with code vector $\vec{b}$ as in the right plot.
						Every $\vec{s} \in \Kb\bb{\Lambda, d}$ is collinear with $\vec{b}$ on the index set $\Lambda$ (indicated by red stems) with $\Lambda$ representing the identities of the cooperating relays.
						Heights of the black dashed stems, indicating values on $\Lambda^{\comp}$, represent contributions from non-cooperating relays and can vary across different vectors in $\Kb\bb{\Lambda, d}$.}
						\label{fig:sample-domain-type-1}
					\end{figure}
				\fi
			\makeatother

			\corollaryname~\ref{cor:cooperative coding unident} and \theoremname~\ref{thm:sparse cooperative coding unident} stated below imply unidentifiability results for certain joint properties of the $\bb{\Lambda', \vec{b}}$ pair, \ie~for certain joint configurations of the cooperating relay index set and the cooperative code employed.
			The unidentifiability results may hold for both $\vec{h} \in \setR^{n}$ (unstructured channel) and $\vec{h} \in \Kb\bb{\Lambda'', n}$ (subspace-sparse channel), even for a large number of cooperating relays (as measured by the cardinality $\card{\Lambda'}$).
			In what follows, $\Lambda_{\ast} + 1$ denotes the Minkowski sum of the sets $\cc{1}$ and $\Lambda_{\ast}$.

			\begin{corollary}
				\label{cor:cooperative coding unident}
				Let $m \geq 3$ and $n \geq 2$ be arbitrary integers.
				For any given index set $\emptyset \neq \Lambda' \subseteq \cc{2,3,\dotsc,m-1}$, let $\vec{b} \neq \vec{0}$, $\mathcal{K} = \Kb\bb{\Lambda',m} \times \setR^{n}$ and define $p \triangleq \card{\Lambda' \bigcup \bb{\Lambda' - 1}}$.
				If $\Lambda' \bigcap \bb{\Lambda' - 1} = \emptyset$, then there exists a set $\mathcal{G}_{\ast} \subseteq \mathcal{K}/\IdR$ of dimension at least $\bb{m+n-p+1}$, such that every $\bb{\vec{g}, \vec{h}} \in \mathcal{G}_{\ast}$ is unidentifiable.
				Otherwise, if $\Lambda' \bigcap \bb{\Lambda' - 1} \neq \emptyset$ then denote by $\emptyset \neq \Lambda_{\ast} \subseteq \cc{1,2,\dots,\card{\Lambda'}}$, the index subset such that $\forall \vec{g} \in \Kb\bb{\Lambda',m}$, $\vec{g}\bb[\big]{\Lambda' \bigcap \bb{\Lambda' - 1}}$ is collinear with $\vec{b}\bb{\Lambda_{\ast}}$.
				If $\vec{b}\bb{\Lambda_{\ast}}$ and $\vec{b}\bb{\Lambda_{\ast} + 1}$ are collinear, then there exists a set $\mathcal{G}_{\ast} \subseteq \mathcal{K}/\IdR$ of dimension at least $\bb{m+n-p}$, such that every $\bb{\vec{g}, \vec{h}} \in \mathcal{G}_{\ast}$ is unidentifiable.
			\end{corollary}

			\corollaryname~\ref{cor:cooperative coding unident} makes a statement for the partially cooperative coded domain $\Kb\bb{\Lambda',m}$ that is essentially analogous to the statement of \corollaryname~\ref{cor:repetition coding unident} for the repetition coded domain $\Kone\bb{\Lambda',m}$, the only technical distinction being a more elaborate specification of the unidentifiable configurations of the pair $\bb{\Lambda', \vec{b}}$ in the former result.
			Some of the technical conditions on the pair $\bb{\Lambda', \vec{b}}$, for generating unidentifiable configurations, are automatically satisfied when $\vec{b} = \vec{1}$ and hence we chose to present this simpler instantiation of \corollaryname~\ref{cor:cooperative coding unident} in the form of \corollaryname~\ref{cor:repetition coding unident} under the rubric of repetition coding in the last subsection.
			In particular, both domains $\Kb\bb{\Lambda',m}$ and $\Kone\bb{\Lambda',m}$ share the same construction for a candidate unidentifiable input signal, provided the premise of \corollaryname~\ref{cor:cooperative coding unident} is satisfied.
			Furthermore, \corollariesname~\ref{cor:mixed unident moderate},~\ref{cor:repetition coding unident} and~\ref{cor:cooperative coding unident} motivate the following categorization of the $\bb{\Lambda, \vec{b}}$ pairs in preparation for the statement of \theoremname~\ref{thm:sparse cooperative coding unident}.
			
			\begin{definition}
				\label{defn:categorization}
				Let $d \geq 3$ be an arbitrary integer, $\emptyset \neq \Lambda \subseteq \cc{2,3,\dotsc,d-1}$ be a given index set and $\vec{b} \in \setR^{\card{\Lambda}}$ be a code vector.
				Let $p \triangleq \card{\Lambda \bigcup \bb{\Lambda - 1}}$.
				The following mutually exclusive (but not exhaustive) categories for the pair $\bb{\Lambda, \vec{b}}$ are defined.
				\begin{enumerate}
					\setcounter{enumi}{-1}
					\item	If $d \geq 5$, $\emptyset \neq \Lambda \subseteq \cc{3,4,\dotsc,d-2}$ and $\vec{b} = \vec{0}$ then $\bb{\Lambda, \vec{b}}$ is of type 0.
					\item	Let $\vec{b} \neq \vec{0}$ and $\Lambda \bigcap \bb{\Lambda - 1} \neq \emptyset$.
							If $\vec{b}$ is collinear with $\vec{1} \in \setR^{\card{\Lambda}}$ then $\bb{\Lambda, \vec{b}}$ is of type 1.
							If $\vec{b}$ is not collinear with $\vec{1} \in \setR^{\card{\Lambda}}$ then let $\emptyset \neq \Lambda_{\ast} \subseteq \cc{1,2,\dots,\card{\Lambda}}$ denote the index subset such that $\forall \vec{w} \in \Kb\bb{\Lambda,d}$, $\vec{w}\bb[\big]{\Lambda \bigcap \bb{\Lambda - 1}}$ is collinear with $\vec{b}\bb{\Lambda_{\ast}}$.
							If $\vec{b}\bb{\Lambda_{\ast}}$ and $\vec{b}\bb{\Lambda_{\ast} + 1}$ are collinear, then $\bb{\Lambda, \vec{b}}$ is of type 1.
					\item	If $\vec{b} \neq \vec{0}$ and $\Lambda \bigcap \bb{\Lambda - 1} = \emptyset$ then $\bb{\Lambda, \vec{b}}$ is of type 2.
				\end{enumerate}
			\end{definition}
			
			For a visual example, we note that \figurename~\ref{fig:sample-K0} represents a $\bb{\Lambda, \vec{b}}$ pair of type 0 while \figurename~\ref{fig:sample-domain-type-1} shows a type 1 pair with $\Lambda_{\ast} = \cc{1,3,4}$ ($\Lambda_{\ast}$ as defined above in \definitionname~\ref{defn:categorization}).
			
			\begin{theorem}
				\label{thm:sparse cooperative coding unident}
				Let $m,n \geq 3$ be arbitrary integers.
				For any given index sets $\emptyset \neq \Lambda' \subseteq \cc{2,3,\dotsc,m-1}$ and $\emptyset \neq \Lambda'' \subseteq \cc{2,3,\dotsc,n-1}$, let $\mathcal{K} = \Kb\bb{\Lambda',m} \times \mathcal{K}_{\vec{b}'}\bb{\Lambda'',n}$ and let $p \triangleq \card{\Lambda' \bigcup \bb{\Lambda' - 1}}$ and $p' \triangleq \card{\Lambda'' \bigcup \bb{\Lambda'' - 1}}$.
				If the pairs $\bb{\Lambda', \vec{b}}$ and $\bb{\Lambda'', \vec{b}'}$ are respectively of types $t$ and $t'$ for $t,t' \in \cc{0,1,2}$, then there exists a set $\mathcal{G}_{\ast} \subseteq \mathcal{K}/\IdR$ of dimension at least $\bb{m+n-1 - p - p' + t + t'}$ such that every $\bb{\vec{g}, \vec{h}} \in \mathcal{G}_{\ast}$ is unidentifiable.
			\end{theorem}

			\begin{IEEEproof}
				\appendixname~\ref{sec:sparse cooperative coding unident proof}.
			\end{IEEEproof}

			\theoremname~\ref{thm:sparse cooperative coding unident} stated above utilizes the construction of candidate adversarial inputs over different domains, as developed in \corollariesname~\ref{cor:mixed unident moderate} and~\ref{cor:cooperative coding unident}, and fuses them together to generate candidate unidentifiable signal pairs by exploiting the separability of the feasible domain $\mathcal{K}$.
			It is straightforward to check that \theoremname~\ref{thm:sparse unident moderate} and \corollariesname~\ref{cor:mixed unident moderate},~\ref{cor:repetition coding unident},~\ref{cor:sparse repetition coding unident}, and~\ref{cor:cooperative coding unident} can all be derived from \theoremname~\ref{thm:sparse cooperative coding unident} by suitable choices of the parameter pairs $\bb{\Lambda', \vec{b}}$ and $\bb{\Lambda'', \vec{b}'}$, \eg~setting $\vec{b} = \vec{1}$ and $\vec{b}' = \vec{0}$ gives back \corollaryname~\ref{cor:sparse repetition coding unident}.
			In essence, barring \corollaryname~\ref{cor:mixed unident everywhere} which requires special assumptions, \theoremname~\ref{thm:sparse cooperative coding unident} encompasses all of the unidentifiability results for constrained blind deconvolution developed in this part of the paper.
			
			The main message of \theoremname~\ref{thm:sparse cooperative coding unident} is that the coded subspaces represented by the vectors $\vec{b}$ and $\vec{b}'$ are critical to the identifiability of the blind linear deconvolution problem~\eqref{prob:find_xy}.
			If these coded subspaces are solely determined by the application or natural system configuration (like the multi-hop channel estimation example above) then sparsity may not be sufficient to guarantee identifiability for the blind deconvolution problem.
			A certain amount of design freedom on \textit{both} of the coded subspaces (represented by vectors $\vec{b}$ and $\vec{b}'$) is \emph{necessary} to guarantee identifiability under blind deconvolution (like in \cite{ahmed2012blind,choudhary2013bilinear,johnson1998blind}).

			As yet another important special case for wireless communications, we interpret the vector $\vec{b}$ as a power-delay profile and assume it to be geometrically decaying over some contiguous index subset $\Lambda$, \ie~$\vec{b}$ satisfies $\vec{b}\bb{j} = r\vec{b}\bb{j-1}$ for some $0 < \abs{r} < 1$ and \emph{all} $j \in \cc{2,3,\dotsc,\card{\Lambda}}$.
			Then we get the \emph{geometrically decaying} cone $\Kb\bb{\Lambda, d}$ of all $d$-dimensional real vectors that are geometrically decaying on the index subset $\Lambda$ and it admits the following unidentifiability result (in the statement below, $\Kb\bb{\Lambda_{2}, n}$ represents a geometrically decaying cone).
			\begin{corollary}
				\label{cor:sparse geometric unident}
				Let $m \geq 3$ and $n \geq 5$ be arbitrary integers.
				For any given index subset $\emptyset \neq \Lambda_{1} \subseteq \cc{3,4,\dotsc,m-2}$ and any contiguous index subset $\emptyset \neq \Lambda_{2} \subseteq \cc{2,3,\dotsc,n-1}$, let $\mathcal{K} = \Kzero\bb{\Lambda_{1},m} \times \Kb\bb{\Lambda_{2},n}$ and define $p_{1} \triangleq \card{\Lambda_{1} \bigcup \bb{\Lambda_{1} - 1}}$, $p_{2} \triangleq \card{\Lambda_{2}} + 1$.
				Then there exists a set $\mathcal{G} \subseteq \mathcal{K}/\IdR$ such that every $\bb{\vec{g}, \vec{h}} \in \mathcal{G}$ is unidentifiable.
				If $\card{\Lambda_{2}} = 1$ then $\mathcal{G}$ is of dimension at least $\bb{m+n+1 - p_{1} - p_{2}}$, otherwise $\mathcal{G}$ is of dimension at least $\bb{m+n - p_{1} - p_{2}}$.
			\end{corollary}
			We omit the proof of \corollaryname~\ref{cor:sparse geometric unident} here as it follows from specializing the proof of \theoremname~\ref{thm:sparse cooperative coding unident}.
			A complete proof of \corollaryname~\ref{cor:sparse geometric unident} appears in our earlier work~\cite{choudhary2014subspaceblind}.
			The occurrence of a geometrically decaying power-delay profile is common in underwater acoustic communication channels~\cite{michelusi2011evaluationhybridsparsediffuse} and more generally in wide-band communication channels~\cite{molisch2006ComprehensiveStandardizedModel,saleh1987StatisticalModelIndoor}.

	\section{Conclusions}
		\label{sec:conclusion}
		Blind deconvolution is an important non-linear inverse problem commonly encountered in signal processing applications.
		Natively, blind deconvolution is ill-posed from the viewpoint of signal identifiability and it is assumed that application specific additional constraints (like sparsity) would suffice to guarantee identifiability for this inverse problem.
		In the current work, we showed that (somewhat surprisingly) sparsity in the canonical basis is insufficient to guarantee identifiability under fairly generic assumptions on the support of the sparse signal.
		Specifically, we explicitly demonstrate a form of rotational ambiguity that holds true for sparse vectors as well.
		Our approach builds on the \emph{lifting} technique from optimization to reformulate blind deconvolution into a rank one matrix recovery problem, and analyzes the rank two null space of the resultant linear operator.
		While this approach is philosophically applicable to other bilinear inverse problems (like dictionary learning), it is the simplicity of the convolution operator in the lifted domain that makes our analysis tractable.
		We developed scaling laws quantifying the ill-posedness of canonical-sparse blind deconvolution by bounding the dimension of the unidentifiable sparse signal set.
		When applied to a second-hop sparse channel estimation problem, our methods also revealed the insufficiency of side information involving repetition coding or geometrically decaying signals towards guaranteeing identifiability under blind linear deconvolution.
		To establish our results, we developed a measure theoretically tight, partially parametric and partially recursive characterization of the rank two null space of the linear convolution map.
		This result is a precursor to non-randomized code design strategies for guaranteeing signal identifiability under the bilinear observation model of linear convolution.
		The design of such codes is a topic of ongoing research.

	\makeatletter
		\ifbool{@bibtex}{
			\bibliographystyle{IEEEtran}
			\bibliography{IEEEabrv,UWA,ownpub,PaperList}
		}{
			\printbibliography[heading=bibintoc]
		}
	\makeatother

	\appendices

	\section{Proof of \theoremname~\ref{thm:sparse unident moderate}}
		\label{sec:sparse unident moderate proof}
		The proof relies on constructing a generative model for vectors $\vec{x} \in \Kzero\bb{\Lambda_{1},m}$ and $\vec{y} \in \Kzero\bb{\Lambda_{2},n}$ such that $\bb{\vec{x}, \vec{y}}$ is unidentifiable within $\mathcal{K} = \Kzero\bb{\Lambda_{1},m} \times \Kzero\bb{\Lambda_{2},n}$.
		Let $\vec{u} \in \Kzero\bb{\Lambda_{1} \bigcup \bb{\Lambda_{1} - 1}, m-1}$, $\vec{v} \in \Kzero\bb{\Lambda_{2} \bigcup \bb{\Lambda_{2} - 1}, n-1}$, and $\bb{\theta,\phi} \in \Gangle \subseteq \setA^{2}$ be chosen arbitrarily, where
		\makeatletter
			\if@twocolumn
				\begin{equation}
					\begin{split}
						\Gangle	& = \mleft\{\bb{\beta, \gamma} \in \setA^{2} \, \middle| \, \beta, \gamma \not \in \set{l\pi/2}{l \in \setZ}, \mright.	\\
						& \qquad \quad \mleft. \vphantom{\setA^{2}}
						\beta - \gamma \not \in \set{l\pi}{l \in \setZ} \mright\}.
					\end{split}
					\label{eqn:angle parameter set}
				\end{equation}
			\else
				\begin{equation}
					\Gangle	= \set{\bb{\beta, \gamma} \in \setA^{2}}{\beta, \gamma \not \in \set{l\pi/2}{l \in \setZ}, \, \beta - \gamma \not \in \set{l\pi}{l \in \setZ}}.
					\label{eqn:angle parameter set}
				\end{equation}
			\fi
		\makeatother
		We use the 4-tuple $\bb{\vec{u}, \vec{v}, \theta, \phi}$ to generate the vector pair $\bb{\vec{x}, \vec{y}}$ using
		\begin{equation}
			\vec{x} =	\begin{bmatrix}
							\vec{u}	&	0	\\
							0	&	-\vec{u}
						\end{bmatrix}
						\begin{bmatrix}
							\cos \theta	\\
							\sin \theta
						\end{bmatrix},
			\quad
			\vec{y} =	\begin{bmatrix}
							0	&	\vec{v}	\\
							\vec{v}	&	0
						\end{bmatrix}
						\begin{bmatrix}
							\sin \phi	\\
							-\cos \phi
						\end{bmatrix},
			\label{eqn:decomposition equation}
		\end{equation}
		and the vector pair $\bb{\vec{x}', \vec{y}'}$ using
		\begin{alignat}{2}
			\vec{x}'	& =	\begin{bmatrix}
								\vec{u}	&	0	\\
								0	&	-\vec{u}
							\end{bmatrix}
							\begin{bmatrix}
								\cos \phi	\\
								\sin \phi
							\end{bmatrix}, %
						& \quad %
			\vec{y}'	& =	\begin{bmatrix}
								0	&	\vec{v}	\\
								\vec{v}	&	0
							\end{bmatrix}
							\begin{bmatrix}
								\sin \theta	\\
								-\cos \theta
							\end{bmatrix}.
			\label{eqn:adversarial pair}
		\end{alignat}
		By assumption, $0 \not \in \cc{\vec{u}(1), \vec{u}(m-1), \vec{v}(1), \vec{v}(n-1)}$ and $0 \not \in \cc{\sin \theta, \cos \theta, \sin \phi, \cos \phi}$, so \eqref{eqn:decomposition equation} implies $0 \not \in \cc{\vec{x}(1), \vec{x}(m), \vec{y}(1), \vec{y}(n)}$ and \eqref{eqn:adversarial pair} implies $0 \not \in \cc{\vec{x}'(1), \vec{x}'(m), \vec{y}'(1), \vec{y}'(n)}$.
		We also have $\vec{u}\bb{\Lambda_{1}} = \vec{u}\bb{\Lambda_{1} - 1} = \vec{0}$ and $\vec{v}\bb{\Lambda_{2}} = \vec{v}\bb{\Lambda_{2} - 1} = \vec{0}$ by assumption, and
		\makeatletter
			\if@twocolumn
				\begin{subequations}
					\label{eqn:zero set pairs}
					\begin{align}
						\begin{split}
							\vec{x}\bb{\Lambda_{1}}	& = \vec{u}\bb{\Lambda_{1}} \cos \theta - \vec{u}\bb{\Lambda_{1} - 1} \sin \theta = \vec{0},	\\
							\vec{y}\bb{\Lambda_{2}}	& = -\vec{v}\bb{\Lambda_{2}} \cos \phi + \vec{v}\bb{\Lambda_{2} - 1} \sin \phi = \vec{0},
						\end{split}
						\label{eqn:pair 1 zero set} \\
						\intertext{as well as}
						\begin{split}
							\vec{x}'\bb{\Lambda_{1}}	& = \vec{u}\bb{\Lambda_{1}} \cos \phi - \vec{u}\bb{\Lambda_{1} - 1} \sin \phi = \vec{0},	\\
							\vec{y}'\bb{\Lambda_{2}}	& = -\vec{v}\bb{\Lambda_{2}} \cos \theta + \vec{v}\bb{\Lambda_{2} - 1} \sin \theta = \vec{0},
						\end{split}
						\label{eqn:pair 2 zero set}
					\end{align}
				\end{subequations}
			\else
				\begin{subequations}
					\label{eqn:zero set pairs}
					\begin{alignat}{2}
						\vec{x}\bb{\Lambda_{1}}	& = \vec{u}\bb{\Lambda_{1}} \cos \theta - \vec{u}\bb{\Lambda_{1} - 1} \sin \theta = \vec{0},	& \quad \vec{y}\bb{\Lambda_{2}}	& = -\vec{v}\bb{\Lambda_{2}} \cos \phi + \vec{v}\bb{\Lambda_{2} - 1} \sin \phi = \vec{0},	\label{eqn:pair 1 zero set} \\
						\vec{x}'\bb{\Lambda_{1}}	& = \vec{u}\bb{\Lambda_{1}} \cos \phi - \vec{u}\bb{\Lambda_{1} - 1} \sin \phi = \vec{0},	& \quad \vec{y}\bb{\Lambda_{2}}	& = -\vec{v}\bb{\Lambda_{2}} \cos \theta + \vec{v}\bb{\Lambda_{2} - 1} \sin \theta = \vec{0},	\label{eqn:pair 2 zero set}
					\end{alignat}
				\end{subequations}
			\fi
		\makeatother
		where \eqref{eqn:pair 1 zero set} follows from \eqref{eqn:decomposition equation} and \eqref{eqn:pair 2 zero set} follows from \eqref{eqn:adversarial pair}.
		Therefore, $\vec{x}, \vec{x}' \in \Kzero\bb{\Lambda_{1},m}$ and $\vec{y}, \vec{y}' \in \Kzero\bb{\Lambda_{2},n}$, implying that $\bb{\vec{x}, \vec{y}}, \bb{\vec{x}', \vec{y}'} \in \mathcal{K}$.
		Setting $\mat{X} = \vec{x}\tpose{\vec{y}} - \vec{x}'\tpose{\bb{\vec{y}'}}$, we have
		\begin{equation}
			\begin{split}
				\mat{X}	& =	\begin{bmatrix}
								\vec{u}	&	0	\\
								0	&	-\vec{u}
							\end{bmatrix}
							\begin{bmatrix}
								\cos \theta	\\
								\sin \theta
							\end{bmatrix}
							\begin{bmatrix}
								\sin \phi	&	-\cos \phi
							\end{bmatrix}
							\begin{bmatrix}
								0		&	\tpose{\vec{v}}	\\
								\tpose{\vec{v}}	&	0
							\end{bmatrix}	\\
						& {} +	\begin{bmatrix}
									\vec{u}	&	0	\\
									0	&	-\vec{u}
								\end{bmatrix}
								\begin{bmatrix}
									\cos \phi	\\
									\sin \phi
								\end{bmatrix}
								\begin{bmatrix}
									-\sin \theta	&	\cos \theta
								\end{bmatrix}
								\begin{bmatrix}
									0		&	\tpose{\vec{v}}	\\
									\tpose{\vec{v}}	&	0
								\end{bmatrix}	\\
						& =	\sin\bb{\phi - \theta}	\begin{bmatrix}
														\vec{u}	&		0	\\
														0	&	-\vec{u}
													\end{bmatrix}
													\begin{bmatrix}
														0	&	\tpose{\vec{v}}	\\
														\tpose{\vec{v}}	&	0
													\end{bmatrix}
			\end{split}
			\label{eqn:adversarial decomposition}
		\end{equation}
		and hence, $\vec{x}\tpose{\vec{y}} - \vec{x}'\tpose{\bb{\vec{y}'}} = \mat{X} \in \mathcal{N}\bb{\mathscr{S}, 2}$ from \lemmaname~\ref{lem:rank-2 nullspace}.
		Therefore, the pairs $\bb{\vec{x}, \vec{y}}$ and $\bb{\vec{x}', \vec{y}'}$ produce the same convolved output and are indistinguishable under the linear convolution map.
		Since $\vec{x}$ and $\vec{x}'$ are linearly independent by the choice $\phi - \theta \not \in \set{l\pi}{l \in \setZ}$, $\bb{\vec{x}, \vec{y}} \in \mathcal{K}$ is unidentifiable by \definitionname~\ref{defn:identifiability} with $\bb{\vec{x}', \vec{y}'}$ as the certificate of unidentifiability.

		All that remains is to lower bound the dimension of a set $\mathcal{G} \subseteq \mathcal{K}$ of pairs $\bb{\vec{x}, \vec{y}}$ that can be shown to be unidentifiable for blind linear deconvolution using the above mentioned construction.
		Let $\mathcal{G}_{\mathrm{V}}$ denote the Cartesian product set $\Kzero\bb{\Lambda_{1} \bigcup \bb{\Lambda_{1} - 1}, m-1} \times \Kzero\bb{\Lambda_{2} \bigcup \bb{\Lambda_{2} - 1}, n-1}$.
		We construct $\mathcal{G}$ as follows.
		A vector pair $\bb{\vec{x}, \vec{y}} \in \mathcal{G}$ if and only if all of the following conditions are satisfied.
		\begin{enumerate}[({A}1)]
			\item	\label{itm:generative condition}
					$\bb{\vec{x}, \vec{y}} \in \setR^{m} \times \setR^{n}$ is generated from a 4-tuple $\bb{\vec{u}, \vec{v}, \theta, \phi} \in \setR^{m-1} \times \setR^{n-1} \times \setA^{2}$ using \eqref{eqn:decomposition equation}.
			\item	\label{itm:angle condition}
					$\bb{\theta, \phi} \in \Gangle$.
			\item	\label{itm:vector condition}
					$\bb{\vec{u}, \vec{v}} \in \mathcal{G}_{\mathrm{V}}$.
			\setcounter{asmplistctr}{\value{enumi}}
		\end{enumerate}
		From the arguments in the previous paragraph, we have $\emptyset \neq \mathcal{G} \subset \Kzero\bb{\Lambda_{1},m} \times \Kzero\bb{\Lambda_{2},n} = \mathcal{K}$ and furthermore, every $\bb{\vec{x}_{\ast}, \vec{y}_{\ast}} \in \mathcal{G}$ is unidentifiable within $\mathcal{K}$.
		Let us consider the sets
		\makeatletter
			\if@twocolumn
				\begin{equation}
					\begin{split}
						\mathcal{G}'_{1}	& = \set{\bb{\vec{u}, \theta} \in \setR^{m-1} \times \setA}{\text{\aref{itm:angle condition} and \aref{itm:vector condition} are true}},	\\
						\mathcal{G}'_{2}\bb{\theta}	& = \mleft\{ \bb{\vec{v}, \phi} \in \setR^{n-1} \times \setA \, \middle| \, \text{\aref{itm:angle condition} and \aref{itm:vector condition} are true}, \mright.	\\
						& \qquad \quad \mleft. \vphantom{\setR^{n-1}}
						\text{given $\theta \in \setA \setminus \set{l\pi/2}{l \in \setZ}$} \mright\}
					\end{split}
					\label{eqn:domain sectioning}
				\end{equation}
			\else
				\begin{equation}
					\begin{split}
						\mathcal{G}'_{1}	& = \set{\bb{\vec{u}, \theta} \in \setR^{m-1} \times \setA}{\text{\aref{itm:angle condition} and \aref{itm:vector condition} are true}},	\\
						\mathcal{G}'_{2}\bb{\theta}	& = \set{\bb{\vec{v}, \phi} \in \setR^{n-1} \times \setA}{\text{given $\theta \in \setA \setminus \set{l\pi/2}{l \in \setZ}$, \aref{itm:angle condition} and \aref{itm:vector condition} are true}}.
					\end{split}
					\label{eqn:domain sectioning}
				\end{equation}
			\fi
		\makeatother
		By the definition in \eqref{eqn:sparse domain}, $\Kzero\bb{\Lambda_{1} \bigcup \bb{\Lambda_{1} - 1}, m-1}$ is a $\bb[\big]{m-1-\card{\Lambda_{1} \bigcup \bb{\Lambda_{1} - 1}}} = \bb{m-1 - p_{1}}$ dimensional Borel subset of $\setR^{m-1}$, and $\Kzero\bb{\Lambda_{2} \bigcup \bb{\Lambda_{2} - 1}, n-1}$ is a $\bb[\big]{n-1-\card{\Lambda_{2} \bigcup \bb{\Lambda_{2} - 1}}} = \bb{n-1 - p_{2}}$ dimensional Borel subset of $\setR^{n-1}$.
		Further, $\setA \setminus \set{l\pi/2}{l \in \setZ}$ is a one dimensional Borel subset of $\setR$ and given a value of $\theta \in \setA$, $\setA \setminus \set{l\pi/2, \theta + l\pi}{l \in \setZ}$ is also a one dimensional Borel subset of $\setR$.
		Hence, \eqref{eqn:domain sectioning}, \aref{itm:angle condition} and \aref{itm:vector condition} imply that
		\begin{enumerate}
			\item	$\mathcal{G}'_{1}$ is a $\bb{m-1 - p_{1}} + 1 = \bb{m - p_{1}}$ dimensional Borel subset of $\setR^{m}$,
			\item	given $\theta \in \setA \setminus \set{l\pi/2}{l \in \setZ}$, $\mathcal{G}'_{2}\bb{\theta}$ is a $\bb{n-1 - p_{2}} + 1 = \bb{n - p_{2}}$ dimensional Borel subset of $\setR^{n}$,
			\item	$\Gangle$ is a two dimensional Borel subset of $\setR^{2}$, and
			\item	$\mathcal{G}_{\mathrm{V}} \times \Gangle$ is a $\bb{m-1 - p_{1}} + \bb{n-1 - p_{2}} + 2 = \bb{m+n - p_{1} - p_{2}}$ dimensional Borel subset of $\setR^{m+n}$.
		\end{enumerate}

		Next, we compute the dimension of the set $\mathcal{G}$.
		Consider a factorization of $\mathcal{G}$ into
		\makeatletter
			\if@twocolumn
				\begin{equation}
					\begin{split}
						\mathcal{G}_{1}	& = \set{\vec{x}_{\ast} \in \setR^{m}}{\bb{\vec{x}_{\ast}, \vec{y}_{\ast}} \in \mathcal{G}},	\\
						\mathcal{G}_{2}\bb{\vec{x}_{\ast}}	& = \set{\vec{y}_{\ast} \in \setR^{n}}{\bb{\vec{x}_{\ast}, \vec{y}_{\ast}} \in \mathcal{G}},
					\end{split}
					\label{eqn:sectioning}
				\end{equation}
			\else
				\begin{equation}
					\mathcal{G}_{1}	= \set{\vec{x}_{\ast} \in \setR^{m}}{\bb{\vec{x}_{\ast}, \vec{y}_{\ast}} \in \mathcal{G}} \quad \text{and} \quad \mathcal{G}_{2}\bb{\vec{x}_{\ast}}	= \set{\vec{y}_{\ast} \in \setR^{n}}{\bb{\vec{x}_{\ast}, \vec{y}_{\ast}} \in \mathcal{G}},
					\label{eqn:sectioning}
				\end{equation}
			\fi
		\makeatother
		\ie~$\mathcal{G} = \set{\bb{\vec{x}_{\ast}, \vec{y}_{\ast}}}{\vec{x}_{\ast} \in \mathcal{G}_{1}, \, \vec{y}_{\ast} \in \mathcal{G}_{2}\bb{\vec{x}_{\ast}}}$.
		Equations \eqref{eqn:sectioning} and \eqref{eqn:decomposition equation} imply that every $\vec{x} \in \mathcal{G}_{1}$ is generated as the result of a uniformly continuous map from $\bb{\vec{u}, \theta} \in \mathcal{G}'_{1}$ (with a non-singular Jacobian matrix).
		From \lemmaname~\ref{lem:finite quotient set}, the quotient set $\mathcal{Q}_{\sim}\bb{\vec{x}, m}$ is finite for every $\vec{x} \in \mathcal{G}_{1}$ and therefore each $\vec{x} \in \mathcal{G}_{1}$ can be generated by at most a finite number of elements $\bb{\vec{u}, \theta} \in \mathcal{G}'_{1}$ using \eqref{eqn:decomposition equation}.
		Given some $\vec{x} \in \mathcal{G}_{1}$, let $\Theta \subset \setA$ be a set such that for every $\theta \in \Theta$, there exists $\bb{\vec{u}, \theta} \in \mathcal{G}'_{1}$ generating $\vec{x}$ using \eqref{eqn:decomposition equation}.
		Clearly, $\card{\Theta} \leq \bb{2m-2}$ from \lemmaname~\ref{lem:finite quotient set}.
		Then, \eqref{eqn:sectioning} and \eqref{eqn:decomposition equation} imply that every $\vec{y} \in \mathcal{G}_{2}\bb{\vec{x}}$ is generated as the result of a uniformly continuous map from $\bb{\vec{v}, \phi} \in \bigcup_{\theta \in \Theta} \mathcal{G}'_{2}\bb{\theta}$ (with a non-singular Jacobian matrix).
		Using \lemmaname~\ref{lem:finite quotient set}, the quotient set $\mathcal{Q}_{\sim}\bb{\vec{y}, n}$ is finite for every $\vec{y} \in \mathcal{G}_{2}\bb{\vec{x}}$ and therefore each $\vec{y} \in \mathcal{G}_{2}\bb{\vec{x}}$ can be generated by at most a finite number of elements $\bb{\vec{v}, \phi} \in \bigcup_{\theta \in \Theta} \mathcal{G}'_{2}\bb{\theta}$ using \eqref{eqn:decomposition equation}.
		Since $\Theta$ is a finite set, the above arguments and \eqref{eqn:sectioning} imply that every element $\bb{\vec{x}, \vec{y}} \in \mathcal{G}$ is generated by at most a finite number of 4-tuples $\bb{\vec{u}, \vec{v}, \theta, \phi} \in \mathcal{G}_{\mathrm{V}} \times \Gangle$, using the uniformly continuous maps in \eqref{eqn:decomposition equation} (with non-singular Jacobian matrices).
		Since $\mathcal{G}_{\mathrm{V}} \times \Gangle$ is a $\bb{m+n - p_{1} - p_{2}}$ dimensional Borel subset of $\setR^{m+n}$, we get $\mathcal{G}$ as a Borel subset of $\setR^{m+n}$ of dimension $\bb{m+n - p_{1} - p_{2}}$.

		We let $\mathcal{G}_{\ast} = \mathcal{G}/\IdR$, \ie~$\mathcal{G}_{\ast}$ is the quotient set of $\mathcal{G}$ \wrt~the equivalence relation $\IdR$.
		Clearly, $\mathcal{G}_{\ast} \subseteq \mathcal{K}/\IdR$.
		Since each element $\bb{\vec{x}, \vec{y}} \in \mathcal{G}/\IdR$ is a representative for a one dimensional Borel subset $\cc{\bb{\alpha \vec{x}, \frac{1}{\alpha} \vec{y}}} \subset \mathcal{G} \subset \setR^{m} \times \setR^{n}$, the dimension of $\mathcal{G}/\IdR$ is one less than the dimension of $\mathcal{G}$.
		Hence, $\mathcal{G}_{\ast}$ is a $\bb{m+n-1 - p_{1} - p_{2}}$ dimensional set.

	\section{Proof of \corollaryname~\ref{cor:mixed unident everywhere}}
		\label{sec:mixed unident everywhere proof}
		This proof follows the template of the proof for \theoremname~\ref{thm:sparse unident moderate} in \appendixname~\ref{sec:sparse unident moderate proof} with adjustments for the change of feasible set $\mathcal{K}$ and borrows from the proof of \theoremname~\ref{thm:ae unident} in \partname~I of the paper to show almost everywhere unidentifiability.
		
		Let $\vec{u} \in \Kzero\bb{\Lambda \bigcup \bb{\Lambda - 1}, m-1}$ and $\vec{y} \in \setR^{n}$ be arbitrarily chosen.
		Since $n \geq 4$ is an even integer, invoking \lemmaname~\ref{lem:finite quotient set} for $\vec{y}$ yields a vector $\vec{v} \in \setR^{n-1}$ and a scalar $\phi \in \setA$ such that the second relationship in \eqref{eqn:decomposition equation} is satisfied.
		Next, we select $\theta \in \setA \setminus \set{l\pi/2}{l \in \setZ}$ and use the parameters $\bb{\vec{u}, \theta}$ in \eqref{eqn:decomposition equation} to generate the vector $\vec{x} \in \setR^{m}$.
		We reuse the parameter 4-tuple $\bb{\vec{u}, \vec{v}, \theta, \phi}$ in \eqref{eqn:adversarial pair} to generate the pair $\bb{\vec{x}', \vec{y}'} \in \setR^{m} \times \setR^{n}$.
		Note that our construction for the 4-tuple $\bb{\vec{u}, \vec{v}, \theta, \phi}$ here is different from that in \appendixname~\ref{sec:sparse unident moderate proof}, even if we set $\Lambda_{2} = \emptyset$ in \theoremname~\ref{thm:sparse unident moderate}.
		Nonetheless, using \eqref{eqn:adversarial decomposition} and reasoning on the same lines as in \appendixname~\ref{sec:sparse unident moderate proof} leads to the conclusion that $\bb{\vec{x}, \vec{y}} \in \mathcal{K}$ is unidentifiable with $\bb{\vec{x}', \vec{y}'}$ as the certificate of unidentifiability.

		For notational brevity, let us define
		\begin{equation}
			\mathcal{H} \triangleq \Kzero\bb[\Big]{\Lambda \bigcup \bb{\Lambda - 1}, m-1} \times \bb[\big]{\setA \setminus \set{l\pi/2}{l \in \setZ}}.
		\end{equation}
		To show that given any $\bb{\vec{u},\theta} \in \mathcal{H}$, almost every choice of $\vec{y} \in \setR^{n}$ yields an unidentifiable pair $\bb{\vec{x}, \vec{y}} \in \mathcal{K}$, it suffices to show that $\phi \not \in \set{l\pi/2, \theta + l\pi}{l \in \setZ}$ holds almost everywhere \wrt~the measure over $\vec{y}$.
		From \eqref{eqn:decomposition equation}, the dependence of $\vec{y}$ is uniformly continuous on the $n$ real numbers $\bb{\phi, \vec{v}(1), \vec{v}(2), \dots, \vec{v}(n-1)}$ and these $n$ real numbers completely parametrize $\vec{y}$.
		Further, since the measure over $\vec{y}$ is absolutely continuous \wrt~the $n$ dimensional Lebesgue measure, it is possible to choose a measure over $\phi$ that is absolutely continuous \wrt~the one dimensional Lebesgue measure.
		Since $\set{l\pi/2, \theta + l\pi}{l \in \setZ}$ is a zero dimensional set (given the value of $\theta$), $\phi \not \in \set{l\pi/2, \theta + l\pi}{l \in \setZ}$ is true almost everywhere \wrt~the measure over $\phi$ and therefore also \wrt~the measure over $\vec{y}$.

		Let $\mathcal{H}' \subseteq \Kzero\bb{\Lambda,m}$ denote the image of $\mathcal{H}$ under the $\bb{\vec{u}, \theta} \mapsto \vec{x}$ map in \eqref{eqn:decomposition equation}.
		To complete the proof, we need to bound the dimension of the set $\mathcal{H}'$.
		Analogous to \appendixname~\ref{sec:sparse unident moderate proof}, we have $\Kzero\bb{\Lambda \bigcup \bb{\Lambda - 1}, m-1}$ as a $\bb{m-1-p}$ dimensional Borel subset of $\setR^{m-1}$ and therefore, $\mathcal{H}$ is a $\bb{m-1-p} + 1 = \bb{m-p}$ dimensional Borel subset of $\setR^{m}$.
		For any $\vec{x} \in \mathcal{H}'$, \lemmaname~\ref{lem:finite quotient set} implies the finiteness of the quotient set $\mathcal{Q}_{\sim}\bb{\vec{x}, m}$ and therefore each $\vec{x} \in \mathcal{H}'$ can be generated by at most a finite number of elements $\bb{\vec{u}, \theta} \in \mathcal{H}$ using \eqref{eqn:decomposition equation}.
		Since \eqref{eqn:decomposition equation} represents uniformly continuous maps with non-singular Jacobian matrices, we have $\mathcal{H}'$ as a $\bb{m-p}$ dimensional Borel subset of $\setR^{m}$.
		The proof is complete since for any $\vec{x} \in \mathcal{H}'$, $\bb{\vec{x}, \vec{y}} \in \mathcal{K}$ is unidentifiable almost everywhere \wrt~any measure over $\vec{y}$ that is absolutely continuous \wrt~the $n$ dimensional Lebesgue measure.

	\section{Proof of \theoremname~\ref{thm:sparse cooperative coding unident}}
		\label{sec:sparse cooperative coding unident proof}
		For notational convenience, we denote $\vec{g}$ by $\vec{x}$, $\vec{h}$ by $\vec{y}$, $\Kb\bb{\Lambda',m}$ by $\mathcal{D}_{1}\bb{m}$, and $\mathcal{K}_{\vec{b}'}\bb{\Lambda'',n}$ by $\mathcal{D}_{2}\bb{n}$.
		The high level architecture of this proof is similar to the proof of \theoremname~\ref{thm:sparse unident moderate} in \appendixname~\ref{sec:sparse unident moderate proof}.
		However, the generative model for unidentifiable pairs $\bb{\vec{x}, \vec{y}} \in \mathcal{K}$ is substantially different, owing to the change in the feasible set $\mathcal{K}$.

		We shall use \eqref{eqn:decomposition equation} to construct unidentifiable pairs $\bb{\vec{x}, \vec{y}}$ in the separable feasible set $\mathcal{K} = \mathcal{D}_{1}\bb{m} \times \mathcal{D}_{2}\bb{n}$.
		Since the generative parameters for $\vec{x}$ and $\vec{y}$ in \eqref{eqn:decomposition equation} are disjoint, an analogous construction also works for the feasible set $\mathcal{K}' = \mathcal{D}_{2}\bb{m} \times \mathcal{D}_{1}\bb{n}$ to produce unidentifiability results.
		For example, \corollaryname~\ref{cor:sparse repetition coding unident} asserts that for $\mathcal{K} = \Kone\bb{\Lambda',m} \times \Kzero\bb{\Lambda'',n}$ and $\Lambda' \bigcap \bb{\Lambda' - 1} \neq \emptyset$, there exists an unidentifiable subset of $\mathcal{K}/\IdR$ of dimension $\bb{m+n - p_{1} - p_{2}}$.
		Our reasoning based on disjoint generation would imply that for $\mathcal{K}' = \Kzero\bb{\Lambda',m} \times \Kone\bb{\Lambda'',n}$ and $\Lambda'' \bigcap \bb{\Lambda'' - 1} \neq \emptyset$, there exists an unidentifiable subset of $\mathcal{K}'/\IdR$ of dimension $\bb{m+n - p_{1} - p_{2}}$.
		Thus, it suffices to consider \textit{unordered} pairs $\bb{t,t'} \in \cc{0,1,2}^{2}$ for the purpose of this proof.
		Note that \theoremname~\ref{thm:sparse unident moderate} is exactly equivalent to the case of $\bb{t,t'} = \bb{0,0}$.
		In the rest of the proof, we complete the treatment for the remaining cases $\bb{t,t'} \in \cc{0,1,2}^{2} \setminus \cc{\bb{0,0}}$.

		Let us choose the 4-tuple of parameters $\bb{\vec{u}, \vec{v}, \theta, \phi} \in \setR^{m-1} \times \setR^{n-1} \times \setA^{2}$ such that we generate $\bb{\vec{x}, \vec{y}} \in \mathcal{K} = \Kb\bb{\Lambda',m} \times \mathcal{K}_{\vec{b}'}\bb{\Lambda'',n}$ from \eqref{eqn:decomposition equation} and $\bb{\vec{x}', \vec{y}'} \in \mathcal{K}$ from \eqref{eqn:adversarial pair}.
		For the construction to be consistent, $\vec{u}$ must satisfy
		\begin{subequations}
			\label{eqn:co-op code ambiguity eqn}
			\begin{align}
			\vec{x}\bb{\Lambda'}	& = \vec{u}\bb{\Lambda'} \cos \theta - \vec{u}\bb{\Lambda' - 1} \sin \theta	= c_{1}\bb{\theta, \phi} \vec{b},	\\
			\vec{x}'\bb{\Lambda'}	& = \vec{u}\bb{\Lambda'} \cos \phi - \vec{u}\bb{\Lambda' - 1} \sin \phi	= c_{2}\bb{\theta, \phi} \vec{b},
			\end{align}
		\end{subequations}
		for some scalars $c_{1}\bb{\theta, \phi}, c_{2}\bb{\theta, \phi} \in \setR \setminus \cc{0}$ that could depend on $\bb{\theta, \phi}$, and $0 \not \in \cc{\vec{u}\bb{1}, \vec{u}\bb{m-1}}$.
		Solving \eqref{eqn:co-op code ambiguity eqn} for $\vec{u}\bb{\Lambda'}$ and $\vec{u}\bb{\Lambda' - 1}$, we get
		\begin{subequations}
			\label{eqn:co-op code generative ambiguity set}
			\begin{align}
				\vec{u}\bb{\Lambda'}	& = \bb{\frac{c_{1}\bb{\theta, \phi} \sin \phi - c_{2}\bb{\theta, \phi} \sin \theta}{\sin\bb{\phi - \theta}}} \vec{b},	\label{eqn:co-op code consistency 1} \\
				\vec{u}\bb{\Lambda' - 1}	& = \bb{\frac{c_{1}\bb{\theta, \phi} \cos \phi - c_{2}\bb{\theta, \phi} \cos \theta}{\sin\bb{\phi - \theta}}} \vec{b}.	\label{eqn:co-op code consistency 2}
			\end{align}
		\end{subequations}
		Analogously, consistency also requires $\vec{v}$ to satisfy
		\begin{subequations}
			\label{eqn:co-op code 2nd ambiguity eqn}
			\begin{align}
				\vec{y}\bb{\Lambda''}	& = \vec{v}\bb{\Lambda'' - 1} \sin \phi - \vec{v}\bb{\Lambda''} \cos \phi	= c'_{1}\bb{\theta, \phi} \vec{b}',	\\
				\vec{y}'\bb{\Lambda''}	& = \vec{v}\bb{\Lambda'' - 1} \sin \theta - \vec{v}\bb{\Lambda''} \cos \theta	= c'_{2}\bb{\theta, \phi} \vec{b}',
			\end{align}
		\end{subequations}
		for some scalars $c'_{1}\bb{\theta, \phi}, c'_{2}\bb{\theta, \phi} \in \setR \setminus \cc{0}$ that could depend on $\bb{\theta, \phi}$, and $0 \not \in \cc{\vec{v}\bb{1}, \vec{v}\bb{n-1}}$.
		Solving \eqref{eqn:co-op code 2nd ambiguity eqn} for $\vec{v}\bb{\Lambda''}$ and $\vec{v}\bb{\Lambda'' - 1}$, we get
		\begin{subequations}
			\label{eqn:co-op code 2nd generative ambiguity set}
			\begin{align}
				\vec{v}\bb{\Lambda''}	& = \bb{\frac{c'_{1}\bb{\theta, \phi} \sin \theta - c'_{2}\bb{\theta, \phi} \sin \phi}{\sin\bb{\phi - \theta}}} \vec{b}',	\label{eqn:co-op code 2nd consistency 1} \\
				\vec{v}\bb{\Lambda'' - 1}	& = \bb{\frac{c'_{1}\bb{\theta, \phi} \cos \theta - c'_{2}\bb{\theta, \phi} \cos \phi}{\sin\bb{\phi - \theta}}} \vec{b}',	\label{eqn:co-op code 2nd consistency 2}
			\end{align}
		\end{subequations}
		Given a value of $\bb{\theta,\phi} \in \Gangle$ ($\Gangle$ is defined in \eqref{eqn:angle parameter set}), if there exist solutions to $\vec{u}$ satisfying \eqref{eqn:co-op code generative ambiguity set} and $\vec{v}$ satisfying \eqref{eqn:co-op code 2nd generative ambiguity set}, then the construction of $\bb{\vec{x}, \vec{y}}, \bb{\vec{x}', \vec{y}'} \in \mathcal{K}$ is consistent.
		Invoking \eqref{eqn:adversarial decomposition} and using the same line of reasoning as in \appendixname~\ref{sec:sparse unident moderate proof} implies that $\bb{\vec{x}, \vec{y}} \in \mathcal{K}$ is unidentifiable with $\bb{\vec{x}', \vec{y}'}$ as the certificate of unidentifiability.
		Furthermore, the consistency condition \eqref{eqn:co-op code generative ambiguity set} for $\vec{u}$ is independent of the consistency condition \eqref{eqn:co-op code 2nd generative ambiguity set} for $\vec{v}$ for a given $\bb{\theta, \phi} \in \Gangle$.
		Thus, it suffices to analyze the cases $t \in \cc{0,1,2}$ for different values of $\bb{\theta, \phi}$ and the same analysis would apply to the cases $t' \in \cc{0,1,2}$ owing to the similarity between \eqref{eqn:co-op code generative ambiguity set} and \eqref{eqn:co-op code 2nd generative ambiguity set}.

		If $\bb{\Lambda', \vec{b}}$ is of type $t = 0$, the analysis in \appendixname~\ref{sec:sparse unident moderate proof} can be reused.
		In particular, we have $\vec{b} = \vec{0}$ and \eqref{eqn:co-op code generative ambiguity set} reduces to the requirement $\vec{u}\bb[\big]{\Lambda' \bigcup \bb{\Lambda' - 1}} = \vec{0}$.
		Clearly, any $\vec{u} \in \Kzero\bb{\Lambda' \bigcup \bb{\Lambda' - 1}, m-1}$ is a valid assignment and $\Kzero\bb{\Lambda' \bigcup \bb{\Lambda' - 1}, m-1}$ is a $\bb{m-1-p+t}$ dimensional Borel subset of $\setR^{m-1}$ for $t = 0$.

		If $\bb{\Lambda', \vec{b}}$ is of type $t = 2$, then $\Lambda' \bigcap \bb{\Lambda' - 1} = \emptyset$ and there are no common variables between the vectors $\vec{u}\bb{\Lambda'}$ and $\vec{u}\bb{\Lambda' - 1}$.
		Thus, any $\theta \not \in \set{\phi + s\pi}{s \in \setZ}$ will yield a consistent assignment for $\vec{u}\bb[\big]{\Lambda' \bigcup \bb{\Lambda' - 1}}$ through \eqref{eqn:co-op code generative ambiguity set} with arbitrary constants $c_{1}\bb{\theta, \phi}$ and $c_{2}\bb{\theta, \phi}$ satisfying $c_{1}\bb{\theta, \phi} \sin \phi \neq c_{2}\bb{\theta, \phi} \sin \theta$ and $c_{1}\bb{\theta, \phi} \cos \phi \neq c_{2}\bb{\theta, \phi} \cos \theta$.
		Clearly, for $\bb{\theta, \phi} \in \Gangle$, there is a two dimensional set of valid choices for the pair $\bb{c_{1}\bb{\theta, \phi}, c_{2}\bb{\theta, \phi}}$ and therefore $\vec{u}\bb{\Lambda'}$ and $\vec{u}\bb{\Lambda'-1}$ lie (independently of each other) on a one dimensional subspace collinear with $\vec{b} \in \setR^{p/2}$, implying that $\vec{u}\bb[\big]{\Lambda' \bigcup \bb{\Lambda' - 1}}$ lies on a two dimensional subspace of $\setR^{p}$.
		Since $\vec{u}\bb{j}$ is unconstrained $\forall j \not \in \Lambda' \bigcup \bb{\Lambda' - 1}$, the set of consistent assignments of $\vec{u}$ is a Borel subset of $\setR^{m-1}$ of dimension at least $2 + \bb{m-1-p} = \bb{m-1-p+t}$.

		Finally, let us analyze the case when $\bb{\Lambda', \vec{b}}$ is of type $t = 1$.
		We have $\Lambda' \bigcap \bb{\Lambda' - 1} \neq \emptyset$, so $\vec{u}\bb[\big]{\Lambda' \bigcap \bb{\Lambda' - 1}}$ must receive consistent assignments from \eqref{eqn:co-op code consistency 1} and \eqref{eqn:co-op code consistency 2}.
		If $\vec{b}$ is collinear with $\vec{1}$, then \eqref{eqn:co-op code generative ambiguity set} is consistent if and only if $\sin\bb{\phi - \theta} \neq 0$ and
		\makeatletter
			\if@twocolumn
				\begin{multline}
					c_{1}\bb{\theta, \phi} \sin \phi - c_{2}\bb{\theta, \phi} \sin \theta	\\
					=	c_{1}\bb{\theta, \phi} \cos \phi - c_{2}\bb{\theta, \phi} \cos \theta
					\neq 0.
					\label{eqn:simplified consistency rep code}
				\end{multline}
			\else
				\begin{equation}
					c_{1}\bb{\theta, \phi} \sin \phi - c_{2}\bb{\theta, \phi} \sin \theta
					=	c_{1}\bb{\theta, \phi} \cos \phi - c_{2}\bb{\theta, \phi} \cos \theta
					\neq 0.
					\label{eqn:simplified consistency rep code}
				\end{equation}
			\fi
		\makeatother
		If $\vec{b}$ is not collinear with $\vec{1}$, then there exists an index subset $\emptyset \neq \Lambda_{\ast} \subseteq \cc{1,2,\dots,\card{\Lambda'}}$ as characterized in \definitionname~\ref{defn:categorization}.
		Further, \eqref{eqn:co-op code consistency 1} and \eqref{eqn:co-op code consistency 2} respectively imply \eqref{eqn:co-op common assignment 1} and \eqref{eqn:co-op common assignment 2} below.
		\makeatletter
			\if@twocolumn
				\begin{subequations}
					\label{eqn:co-op common assignment}
					\begin{align}
						\vec{u}\bb[\big]{\Lambda' \bigcap \bb{\Lambda' - 1}}	& = \bb{\frac{c_{1}\bb{\theta, \phi} \sin \phi - c_{2}\bb{\theta, \phi} \sin \theta}{\sin\bb{\phi - \theta}}} \vec{b}\bb{\Lambda_{\ast}},	\label{eqn:co-op common assignment 1} \\
						\begin{split}
							\vec{u}\bb[\big]{\Lambda' \bigcap \bb{\Lambda' - 1}}	& {} \\
							& \mspace{-90mu} = \bb{\frac{c_{1}\bb{\theta, \phi} \cos \phi - c_{2}\bb{\theta, \phi} \cos \theta}{\sin\bb{\phi - \theta}}} \vec{b}\bb{\Lambda_{\ast} + 1}.
						\end{split}	\label{eqn:co-op common assignment 2}
					\end{align}
				\end{subequations}
			\else
				\begin{subequations}
					\label{eqn:co-op common assignment}
					\begin{align}
						\vec{u}\bb[\big]{\Lambda' \bigcap \bb{\Lambda' - 1}}	& = \bb{\frac{c_{1}\bb{\theta, \phi} \sin \phi - c_{2}\bb{\theta, \phi} \sin \theta}{\sin\bb{\phi - \theta}}} \vec{b}\bb{\Lambda_{\ast}},	\label{eqn:co-op common assignment 1} \\
						\vec{u}\bb[\big]{\Lambda' \bigcap \bb{\Lambda' - 1}}	& = \bb{\frac{c_{1}\bb{\theta, \phi} \cos \phi - c_{2}\bb{\theta, \phi} \cos \theta}{\sin\bb{\phi - \theta}}} \vec{b}\bb{\Lambda_{\ast} + 1}.	\label{eqn:co-op common assignment 2}
					\end{align}
				\end{subequations}
			\fi
		\makeatother
		Since $\bb{\Lambda', \vec{b}}$ is of type $t = 1$, $\vec{b}\bb{\Lambda_{\ast}}$ and $\vec{b}\bb{\Lambda_{\ast} + 1}$ are collinear and there exists a constant $r \in \setR \setminus \cc{0}$ such that $\vec{b}\bb{\Lambda_{\ast} + 1} = r\vec{b}\bb{\Lambda_{\ast}}$.
		The consistency requirement in \eqref{eqn:co-op common assignment} then reduces to requiring $\sin\bb{\phi - \theta} \neq 0$ and
		\makeatletter
			\if@twocolumn
				\begin{multline}
					c_{1}\bb{\theta, \phi} \sin \phi - c_{2}\bb{\theta, \phi} \sin \theta	\\
					= r \cdot c_{1}\bb{\theta, \phi} \cos \phi - r \cdot c_{2}\bb{\theta, \phi} \cos \theta
					\neq 0.
					\label{eqn:consistency intermediate step}
				\end{multline}
			\else
				\begin{equation}
					c_{1}\bb{\theta, \phi} \sin \phi - c_{2}\bb{\theta, \phi} \sin \theta
					=	r \cdot c_{1}\bb{\theta, \phi} \cos \phi - r \cdot c_{2}\bb{\theta, \phi} \cos \theta
					\neq 0.
					\label{eqn:consistency intermediate step}
				\end{equation}
			\fi
		\makeatother
		Note that \eqref{eqn:consistency intermediate step} subsumes \eqref{eqn:simplified consistency rep code} (on setting $r = 1$), and further implies
		\begin{equation}
			\frac{c_{2}\bb{\theta, \phi}}{c_{1}\bb{\theta, \phi}}
			= \frac{\sin \phi - r\cos \phi}{\sin \theta - r\cos \theta}
			= \frac{\sin \bb{\phi - B}}{\sin \bb{\theta - B}}
			\label{eqn:simplified consistency co-op code}
		\end{equation}
		for $B = \arctan \bb{r}$.
		We define the set $ \Gangle'\bb{r} \subseteq \setA^{2}$ as
		\makeatletter
			\if@twocolumn
				\begin{equation}
					\begin{split}
						 \Gangle'\bb{r}	& = \mleft\{\bb{\beta, \gamma} \in \setA^{2} \, \middle| \, \beta - \gamma \not \in \set{l\pi}{l \in \setZ}, \mright.	\\
						& \qquad \qquad \mleft. \vphantom{\setA^{2}}
						\beta, \gamma \not \in \set{l\pi/2, l\pi+B}{l \in \setZ} \mright\}.
					\end{split}
					\label{eqn:angle parameter set 2}
				\end{equation}
			\else
				\begin{equation}
					 \Gangle'\bb{r}	= \set{\bb{\beta, \gamma} \in \setA^{2}}{\beta, \gamma \not \in \set{l\pi/2, l\pi+B}{l \in \setZ}, \, \beta - \gamma \not \in \set{l\pi}{l \in \setZ}}.
					\label{eqn:angle parameter set 2}
				\end{equation}
			\fi
		\makeatother
		and notice that for any $\bb{\theta, \phi} \in  \Gangle'\bb{r}$, $\sin\bb{\phi - \theta} \neq 0$ and \eqref{eqn:simplified consistency co-op code} implies that the set of valid choices for the pair $\bb{c_{1}\bb{\theta, \phi}, c_{2}\bb{\theta, \phi}}$ is one dimensional.
		Therefore, it follows from \eqref{eqn:co-op code generative ambiguity set} and \eqref{eqn:co-op common assignment} that $\vec{u}\bb[\big]{\Lambda' \bigcup \bb{\Lambda' - 1}}$ lies on a one dimensional subspace of $\setR^{p}$ and is given by
		\begin{equation}
			\begin{bmatrix}
				\vec{u}\bb[\big]{\Lambda' \setminus \bb{\Lambda' - 1}}	\\
				\vec{u}\bb[\big]{\Lambda' \bigcap \bb{\Lambda' - 1}}	\\
				\vec{u}\bb[\big]{\bb{\Lambda' - 1} \setminus \Lambda'}
			\end{bmatrix}
			=	c\bb{\theta, \phi}	\begin{bmatrix}
										\vec{b}\bb{\Lambda_{\ast}^{\comp}}	\\
										\vec{b}\bb{\Lambda_{\ast}}	\\
										\frac{1}{r}\vec{b}\bb{\Lambda_{\ast}^{\comp}}
									\end{bmatrix}
			\label{eqn:1D subspace}
		\end{equation}
		for some scalar $c\bb{\theta, \phi} \in \setR \setminus \cc{0}$.
		In the special case when $\vec{b}$ is collinear with $\vec{1}$, \eqref{eqn:1D subspace} simplifies to $\vec{u}\bb[\big]{\Lambda' \bigcup \bb{\Lambda' - 1}}$ being collinear with $\vec{1}$.
		Since $\vec{u}\bb{j}$ is unconstrained $\forall j \not \in \Lambda' \bigcup \bb{\Lambda' - 1}$, the set of consistent assignments of $\vec{u}$ is a Borel subset of $\setR^{m-1}$ of dimension at least $1 + \bb{m-1-p} = \bb{m-1-p+t}$, for any $\bb{\theta, \phi} \in  \Gangle'\bb{r}$.

		For notational brevity, we define the following.
		Let $\mathcal{G}_{1} \subseteq \setR^{m-1}$ (respectively $\mathcal{G}_{2} \subseteq \setR^{n-1}$) denote the set of all vectors $\vec{u}$ (respectively $\vec{v}$) that satisfy the consistency condition \eqref{eqn:co-op code generative ambiguity set} and $0 \not \in \cc{\vec{u}\bb{1}, \vec{u}\bb{m-1}}$ (respectively \eqref{eqn:co-op code 2nd generative ambiguity set} and $0 \not \in \cc{\vec{v}\bb{1}, \vec{v}\bb{n-1}}$) for a given value of $\bb{\theta, \phi}$.
		Moreover, let $\mathcal{G}_{3} \subseteq \setR^{2}$ denote the largest set of allowed values for the pair $\bb{\theta, \phi}$ depending on the type pair $\bb{t,t'}$ of the pairs $\bb{\Lambda', \vec{b}}$ and $\bb{\Lambda'', \vec{b}'}$.
		In particular,
		\begin{enumerate}
			\item	$t,t' \in \cc{0,2}$ implies $\mathcal{G}_{3} \subseteq \Gangle$,
			\item	$t = 1$ implies $\mathcal{G} \subseteq \Gangle'\bb{r} \subseteq \Gangle$, and
			\item	$t' = 1$ implies $\mathcal{G} \subseteq \Gangle'\bb{r'} \subseteq \Gangle$, where $r'$ depends on $\bb{\Lambda'', \vec{b}'}$ in the same way as $r$ depends on $\bb{\Lambda', \vec{b}}$.
		\end{enumerate}
		Finally, let $\mathcal{G}' \subseteq \mathcal{K}$ denote the image of the set $\mathcal{G}_{1} \times \mathcal{G}_{2} \times \mathcal{G}_{3}$ under the map $\bb{\vec{u}, \vec{v}, \theta, \phi} \mapsto \bb{\vec{x}, \vec{y}}$ in \eqref{eqn:decomposition equation}.
		Clearly, $\mathcal{G}' \neq \emptyset$ and every $\bb{\vec{x}_{\ast}, \vec{y}_{\ast}} \in \mathcal{G}'$ is unidentifiable within $\mathcal{K}$ if \eqref{eqn:co-op code generative ambiguity set} and \eqref{eqn:co-op code 2nd generative ambiguity set} can be simultaneously satisfied with $0 \not \in \cc{\vec{u}\bb{1}, \vec{u}\bb{m-1}, \vec{v}\bb{1}, \vec{v}\bb{n-1}, \sin\phi, \cos\phi, \sin\theta, \cos\theta}$.
		Our analysis indeed guarantees this using \definitionname~\ref{defn:categorization} and careful descriptions for the sets $\mathcal{G}_{1}$, $\mathcal{G}_{2}$, $\mathcal{G}_{3}$ and the conic family in \eqref{eqn:cooperative code defn}.
		
		To finish the proof, we need to compute the dimension of $\mathcal{G}'$.
		Our analysis for the cases $t \in \cc{0,1,2}$ establishes that $\mathcal{G}_{1}$ is a $\bb{m-1-p+t}$ dimensional Borel set and (by analogy) $\mathcal{G}_{2}$ is a $\bb{n-1-p'+t'}$ dimensional Borel set.
		Clearly, $ \Gangle'\bb{r}$ is a two dimensional Borel set for any $r \in \setR$.
		Therefore, $\mathcal{G}_{1} \times \mathcal{G}_{2} \times \mathcal{G}_{3}$ is a $\bb{m-1-p+t} + \bb{n-1-p'+t'} + 2 = \bb{m+n-p-p'+t+t'}$ dimensional Borel subset of $\setR^{m+n}$.
		Since $\mathcal{G}'$ is the concatenated image of the uniformly continuous maps in \eqref{eqn:decomposition equation} (with non-singular Jacobian matrices), invoking \lemmaname~\ref{lem:finite quotient set} as in \appendixname~\ref{sec:sparse unident moderate proof} implies that $\mathcal{G}'$ is a $\bb{m+n-p-p'+t+t'}$ dimensional Borel set.
		Setting $\mathcal{G}_{\ast} \subseteq \mathcal{G}'/\IdR \subseteq \mathcal{K}/\IdR$ and invoking the argument in the last paragraph of \appendixname~\ref{sec:sparse unident moderate proof} implies that the dimension of $\mathcal{G}_{\ast}$ is one less than the dimension of $\mathcal{G}'$ and completes the proof.

\end{document}